\documentclass[aps,prl,amsmath,amssymb,reprint,authoryear,numbers,sort&compress,superscriptaddress,twocolumn]{revtex4-1}
\usepackage{chemfig}
\usepackage{amsmath}
\usepackage{amssymb}
\usepackage{color}
\usepackage{times}
\usepackage{epsf}
\usepackage{graphicx}% Include figure files
\usepackage{epsfig}
\usepackage{epstopdf}
\usepackage{dcolumn}% Align table columns on decimal point
\usepackage{bm}% bold math
\usepackage{float}
\usepackage{url}
\usepackage{CJK}
\usepackage{ulem}
\usepackage{makecell}
\usepackage{newtxtext, newtxmath}
\usepackage{multirow}
\usepackage{subfigure}

\def\kB{k_{\rm B}}
\def\<{\langle}
\def\>{\rangle}
\def\({\left(}
\def\){\right)}
\def\[{\left[}
\def\]{\right]}

\def\kk{\textbf{\textit{k}}}

\def\ss{\textbf{\textit{s}}}

\begin{document}

\title{Abnormal Phonon Angular Momentum due to Off-diagonal Elements in Density Matrix induced by Temperature Gradient}
\author{Jinxin Zhong}
\thanks{These authors contribute equally to this work.}
\affiliation{School of Physics Science and Engineering, Tongji University, 
Shanghai 200092, China}

\author{Hong Sun}
\thanks{These authors contribute equally to this work.}
\affiliation{Phonon Engineering Research Center of Jiangsu Province,
  Center for Quantum Transport and Thermal Energy Science, Institute
  of Physics and Interdisciplinary Science, School of Physics and Technology, Nanjing Normal
	University, Nanjing 210023, China}

\author{Yang Pan}
\affiliation{Phonon Engineering Research Center of Jiangsu Province,
  Center for Quantum Transport and Thermal Energy Science, Institute
  of Physics and Interdisciplinary Science, School of Physics and Technology, Nanjing Normal
	University, Nanjing 210023, China}

\author{Zhiguo Wang}
\affiliation{School of Physics Science and Engineering, Tongji University, 
Shanghai 200092, China}

\author{Xiangfan Xu}
\affiliation{School of Physics Science and Engineering, Tongji University, 
Shanghai 200092, China}

\author{Lifa Zhang}
\email{phyzlf@njnu.edu.cn}
\affiliation{Phonon Engineering Research Center of Jiangsu Province,
  Center for Quantum Transport and Thermal Energy Science, Institute
  of Physics and Interdisciplinary Science, School of Physics and Technology, Nanjing Normal
	University, Nanjing 210023, China}

\author{Jun Zhou}
\email{zhoujunzhou@njnu.edu.cn}
\affiliation{Phonon Engineering Research Center of Jiangsu Province,
  Center for Quantum Transport and Thermal Energy Science, Institute
  of Physics and Interdisciplinary Science, School of Physics and Technology, Nanjing Normal
	University, Nanjing 210023, China}

\date{\today}

\begin{abstract}
Nonzero mean value of phonon angular momentum (PAM) in chiral materials can be generated
when a temperature gradient is applied. We find that both diagonal and
off-diagonal terms of PAM contribute to mean PAM by using the Kubo
formula where both diagonal and off-diagonal elements of the heat
current operator are considered.
The calculation results show that the off-diagonal term is dominant
when the phonon scattering is strong enough.
This finding reveals that the quantum transition between different phonon modes induced by temperature gradient strongly affects the
local atomic rotation. Our discovery provides an explanation of the
recently observed chiral phonon activated spin Seebeck effect.
    
\end{abstract}
%\PACS{asaa}

\maketitle

Phonons are the quanta of atomic displacement field in solids
\cite{Kittelbook}. For a given monatomic lattice, the displacement of
each atom can be obtained through solving the equations of motion once the dynamical
matrix ${\bf D}({\bf k})$ is known where ${\bf k}$ is the phonon wave
vector. When the lattice has inversion symmetry, ${\bf D}({\bf k})$
is a real matrix and ${\bf D}({\bf k})={\bf D}(-{\bf
  k})$. Consequently, the polarization vectors ${\bm \epsilon}_{{\bf k}\sigma}$ must
be real, where $\sigma$ is the branch index, and the system contains only pure vibration. In contrast, when
the inversion symmetry is broken, ${\bf D}({\bf k})$ is a
complex matrix and ${\bm \epsilon}_{{\bf k}\sigma}$ could be
complex. The system does not only contain vibration but also rotation.
%It is straightforward that the time reversal symmetry $D_{\alpha\beta}(-{\bf
%  k})=D_{\alpha\beta}^{\ast}({\bf k})$ and ${\bm
%  \epsilon}_{-{\bf k}\sigma}={\bm \epsilon}_{{\bf k}\sigma}^{\ast}$
%where $\alpha,\beta=x,y,z$ denote directions.

Classically, the displacement of atom in unit cell $l$ at ${\bf
  R}_{l}$ is ${\bf
  u}_{l}\sim{\text {Re}}[{\bm \epsilon}_{{\bf k}\sigma}e^{i({\bf
k}\cdot {\bf R}_{l}-\omega_{{\bf k}\sigma}t)}]$ for given state
$(\sigma,{\bf k})$ \cite{Kittelbook} where $\omega_{{\bf k}\sigma}$ is the phonon frequency. For
real polarization vector, ${\bf u}_{l}\sim{\bm
  \epsilon}_{{\bf k}\sigma}{\text {cos}}({\bf
k}\cdot {\bf R}_{l}-\omega_{{\bf k}\sigma}t)$. The motion of
each atom can be regarded as three ``{\sl in-phase}'' harmonic oscillators which
gives rise to zero angular momentum because ${\bf
  u}_{l}$ is always parallel to $\dot{\bf u}_{l}$.
The case is different for complex polarization vector, the
displacement along $\mu$-direction ($\mu=x,y,z$) is 
\begin{eqnarray}
u_{l}^{\mu}\sim &&{\text
  {Re}}(\epsilon_{{\bf k}\sigma}^{\mu}){\text {cos}}({\bf
k}\cdot {\bf R}_{l}-\omega_{{\bf k}\sigma}t) -{\text
  {Im}}(\epsilon_{{\bf k}\sigma}^{\mu}){\text {sin}}({\bf
k}\cdot {\bf R}_{l}-\omega_{{\bf k}\sigma}t)\nonumber \\
=&&{\text {cos}}({\bf
k}\cdot {\bf R}_{l}-\omega_{{\bf k}\sigma}t-\phi^{\mu}_{{\bf k}\sigma}),
\label{displacement}
\end{eqnarray}
where the phase shift is determined by $\tan(\phi^{\mu}_{{\bf k}\sigma})={\text
    {Im}}(\epsilon_{{\bf k}\sigma}^{\mu})/{\text{Re}}(\epsilon_{{\bf k}\sigma}^{\mu})$.
Eq. (\ref{displacement}) shows that the motion of atom can be
regarded as three ``{\sl out-of-phase}'' harmonic oscillators, because
$\phi^{\mu}_{{\bf k}\sigma}\ne \phi^{\nu}_{{\bf
    k}\sigma}$ when $\mu\ne \nu$. In other words, atoms
rotate around their equilibrium positions circularly or elliptically.
Such microscopic local rotation gives rise to nonzero
angular momentum, which has been experimentally observed in WSe$_2$ \cite{Zhu2018SC}, is termed as phonon angular momentum (PAM) \cite{Zhang2014PRL,Zhang2015PRL}. 

A quantum mechanical theory of PAM was first given by McLellan \cite{Mclellan1988JPC}. The overall PAM of a lattice with $n$ atoms in each unit cell can be written as \cite{Mclellan1988JPC}: 
\begin{equation}
{\bm L}=\sum_{l\kappa}{\bf u}_{l\kappa}\times (m_{\kappa}\dot{\bf
  u}_{l\kappa}).
\label{eq_L}
\end{equation}
$m_{\kappa}$ is the mass of $\kappa$-th atom with $\kappa=1,2,...n$.
In 2014, Zhang and Niu
\cite{Zhang2014PRL} presented a comprehensive second 
quantization form of PAM when the inversion symmetry is absent. 
They found that, when the system is in equilibrium and has time-reversal symmetry,
mean PAM vanishes.
Nonzero mean PAM can be obtained by two possible ways: (1) breaking the time-reversal symmetry \cite{Zhang2014PRL}; (2) driving the
system into non-equilibrium \cite{Hamada2018PRL,Park2020NL,Huang2020ACSNano}.
Later on Hamada {\sl et al.} \cite{Hamada2018PRL} found
a nonzero PAM by using the Boltzmann
transport equation under relaxation time approximation when a
temperature gradient was applied.
% gives $f_{{\bf
%    k}\sigma}=f_{{\bf k}\sigma}^{0}-\tau \frac{\partial
%  f^{0}}{\partial T} {\bf v}_{{\bf
%    k}\sigma}\cdot \nabla T$. 
The $\mu$-component of mean PAM is calculated as $\langle L_{\mu}\rangle=\Lambda_{\mu\nu}\partial T/\partial x_{\nu}$ where
$\Lambda_{\mu\nu}$ is a response tensor. 
%${\bf v}_{{\bf
%    k}\sigma}$ is the phonon group
%velocity and $\tau$ is the phonon relaxation time. 
As a result, a nonzero
phonon magnetic moment due to PAM was calculated accordingly
\cite{Hamada2018PRL,Park2020NL,Ren2021PRL}.

However, the aforementioned calculation results are too small to explain the
recent observed chiral phonon activated Seebeck effect
\cite{Kim2022NM}. A significant magnetic effect was measured by
the time-resolved magneto-optical Kerr effect (TR MOKE) in
chiral organic-inorganic hybrid perovskite with laser pulse
heating. Moreover, the current theory is not able to explain the
observed large magnetic moment in Dirac semimetal Cd$_3$As$_2$ \cite{Chen2020NL}
and ErFeO$_{3}$ \cite{Nova2017NP}. Therefore, there must be other mechanism beyond Hamada's calculation \cite{Hamada2018PRL} which can leads to larger mean PAM and phonon magnetic moment.

In this Letter, we revisit the derivation from Eq. (\ref{eq_L}) and propose a new mechanism to generate nonzero mean PAM by 
keeping both diagonal and off-diagonal terms of density matrix in Kubo formula.
We find that the off-diagonal terms of PAM, which describe the quantum transition between different phonon states, result in notable mean PAM only when the system is not in equilibrium.

We start our study from the $\mu$-component of Eq. (\ref{eq_L}) which
can be written as \cite{HamadaThesis}:
\begin{widetext}
\begin{equation}
	\begin{aligned}
		L_{\mu}\approx\dfrac{\hbar}{2N}\sum\limits_{l}
                \sum\limits_{{\bf k},{\bf k}^{\prime}}
                \sum\limits_{\sigma,\sigma^{\prime}} \(\epsilon_{{\bf
                    k}\sigma}^{\dagger}M_{\mu}\epsilon_{{\bf
                    k}^{\prime}\sigma^{\prime}}\sqrt{\dfrac{\omega_{{\bf
                        k}^{\prime}\sigma^{\prime}}}{\omega_{{\bf
                        k}\sigma}}}a_{{\bf k}\sigma}^{\dagger}a_{{\bf
                    k}^{\prime}\sigma^{\prime}}- \epsilon_{{\bf
                    k}^{\prime}\sigma^{\prime}}^{T}M_{\mu}\epsilon_{{\bf
                    k}\sigma}^{*}\sqrt{\frac{\omega_{{\bf
                        k}\sigma}}{\omega_{{\bf
                        {k}^{\prime}\sigma^{\prime}}}}}
                   a_{{\bf k}^{\prime}\sigma^{\prime}}a_{{\bf
                      k}\sigma}^{\dagger} \) e^{i{\bf R}_{l}\cdot
                    ({\bf k}^{\prime}-{\bf k})}.
		\label{eq17}
	\end{aligned}
\end{equation}
\end{widetext}
$a_{{\bf k}\sigma}^{\dagger}$ and $a_{{\bf k}\sigma}$ are the creation
and annihilation operators of phonons, respectively. Both $aa$ and
$a^{\dagger}a^{\dagger}$ terms are neglected since they 
vary rapidly with time and have marginal contribution.
When the system is a crystal, Eq. (\ref{eq17}) can be further
simplified by using $\frac{1}{N} \sum_{l} e^{i{\bf R}_{l}\cdot ({\bf k}^{\prime}-{\bf k})}=\delta_{{\bf k},{\bf k}^{\prime}}$ and
$-\epsilon_{{\bf k}^{\prime}\sigma^{\prime}}^{T}M_{\mu}\epsilon_{{\bf k}\sigma}^{*}=\epsilon_{{\bf k}\sigma}^{\dagger}M_{\mu}\epsilon_{{\bf k}^{\prime}\sigma^{\prime}}$.
Then $L_{\mu}$ can be divided into two parts: the diagonal term
($L_{\mu}^{\text D}$) and off-diagonal term ($L_{\mu}^{\text {OD}}$) which can
be written as

\begin{subequations}
\begin{equation}
L_{\mu}^{\text D}=\sum\limits_{{\bf k}\sigma}l_{{\bf k}\sigma}^{\mu}\[a_{{\bf k}\sigma}^{\dagger}a_{{\bf k}\sigma}+\frac{1}{2}\],\label{eq_LZD}
\end{equation}
\begin{equation}
L_{\mu}^{\text {OD}}=\sum\limits_{\bf k}
\sum^{\sigma\neq\sigma^{\prime}}\limits_{\sigma\sigma^{\prime}}l^{\mu}_{{\bf k}\sigma\sigma^{\prime}}
 a_{{\bf k}\sigma}^{\dagger}a_{{\bf k}\sigma^{\prime}}.
\label{eq_LZOD}
\end{equation}
\end{subequations}
We note the matrix elements in Eqs. (\ref{eq_LZD}) and (\ref{eq_LZOD})
as follows:
\begin{subequations}
\begin{equation}
l_{{\bf k}\sigma}^{\mu}=\hbar(\epsilon_{{\bf
      k}\sigma}^{\dagger}M_{\mu}\epsilon_{{\bf k}\sigma}),   
\label{eq_MED}
\end{equation}
\begin{equation}
l^{\mu}_{{\bf k}\sigma\sigma^{\prime}}=\frac{\hbar}{2}\epsilon_{{\bf k}\sigma}^{\dagger}M_{\mu}\epsilon_{{\bf
    k}\sigma^{\prime}}\(\sqrt{\frac{\omega_{{\bf
        k}\sigma^{\prime}}}{\omega_{{\bf k}\sigma}}}+\sqrt{\frac{\omega_{{\bf k}\sigma}}{\omega_{{\bf
        k}\sigma^{\prime}}}}\), \space (\sigma\ne\sigma^{\prime}). \label{eq_MEOD}
\end{equation}
\end{subequations}
where $M_{\mu} = I_{n\times n} \bigotimes 
(-i)\varepsilon_{\mu\nu\gamma}$ ($\varepsilon$ is Levi-Civita tensor) and $\hbar$ is the Planck constant. The diagonal term
in Eq. (\ref{eq_LZD}) and its matrix elements in Eq. (\ref{eq_MED})
have been studied in our previous work \cite{Zhang2014PRL}. The
off-diagonal term in Eq.(\ref{eq_LZOD}) and its matrix elements in
Eq. (\ref{eq_MEOD}) show that the off-diagonal PAM describes the quantum transition between state ($\sigma,{\bf k}$) and state ($\sigma^{\prime},{\bf k}$).

We modify the Kubo formula of thermal conductivity \cite{kubo1957statisticalT,zwanzig1965time,mori1962j},
which is similar to the electrical Kubo formula \cite{kubo1957statisticalE}, to
calculate frequency-dependent mean PAM as:
\begin{eqnarray}
	\<L_{\mu}(\omega)\>&=&-V
        \sum\limits_{\nu}\dfrac{(\nabla T)_{\nu}}{T}\sum\limits_{n}\dfrac{e^{-\beta E_{n}}}{Z}\int_{0}^{\beta}d\lambda\int_{0}^{\infty}dte^{i(\omega+i\eta)t}\nonumber\\
        &&\times\<S_{\nu}(-i\hbar\lambda)L_{\mu}(t)\>. 
\label{eq_Kubo}
\end{eqnarray}
Here $T$ is temperature, $\nabla T$ is temperature gradient, $V$ is
volume, $Z$ is the partition function, $\beta=1/k_{B}T$ with $k_{B}$
the Boltzmann constant. $\lambda$ is a parameter. Detailed derivation of Eq. (\ref{eq_Kubo})
is shown in Appendix A. 
The heat current density operator (${\bf S}$) in
Eq.(\ref{eq_Kubo}) is also written into diagonal part (${\bf S}^{\text
  D}$) and off-diagonal part (${\bf S}^{\text {OD}}$) as follows \cite{Allen1993PRB,Hardy1963PR}:

\begin{subequations}
\begin{equation}
{\bf S}^{\text D}=\sum_{{\bf k}\sigma}{\bf s}_{\bf k\sigma}\[a_{{\bf k}\sigma}^{\dagger}a_{{\bf
    k}\sigma}+\frac{1}{2}\], 
\label{eq_HCDOD}
\end{equation}
\begin{equation}
{\bf S}^{\text {OD}}=\sum_{{\bf
    k}}\sum^{\sigma\ne \sigma^{\prime}}_{\sigma\sigma^{\prime}}{\bf s}_{\bf k\sigma\sigma^{\prime}}a_{{\bf k}\sigma}^{\dagger}a_{{\bf
    k}\sigma^{\prime}}.%e^{i(\omega_{{\bf k}\sigma}-\omega_{{\bf k}\sigma^{\prime}})t}
\label{eq_HCDOOD}
\end{equation}
\end{subequations}
The matrix elements are
\begin{subequations}
\begin{equation}
	{\bf s}_{{\bf k}\sigma}=\dfrac{\hbar}{2V}\widetilde\epsilon_{{\bf k}\sigma}^{\hspace{0.1cm}\dagger}                          \dfrac{\partial{\widetilde{\text D}(\bf k)}}{\partial{\bf k}}
\widetilde\epsilon_{{\bf k}{\sigma}},
	 \label{eq_SD}
\end{equation}
\begin{equation}
	{\bf s}_{{\bf k}\sigma\sigma^{\prime}}=\dfrac{\hbar}{4V}\widetilde\epsilon_{{\bf k}\sigma}^{\hspace{0.1cm}\dagger}
\dfrac{\partial{\widetilde{\text D}(\bf k)}}{\partial{\bf k}}
\widetilde\epsilon_{{\bf k}\sigma^{\prime}}
\(\sqrt{\dfrac{\omega_{{\bf k}{\sigma^{\prime}}}}{\omega_{{\bf k}{\sigma}}}}+\sqrt{\dfrac{\omega_{{\bf k}{\sigma}}}{\omega_{{\bf k}{\sigma^{\prime}}}}}\),
\label{eq_SOD}
\end{equation}
\end{subequations}
where $\widetilde{\text D}(\bf k)$ is the redefined dynamical matrix and $\widetilde\epsilon_{{\bf k}\sigma}$ is the corresponding eigenstate. Detail derivations of above terms are given in Appendix B. It has been
proved that ${\bf s}_{\bf k\sigma}=\frac{1}{V}\hbar\omega_{{\bf
    k}\sigma}{\bf v}_{{\bf k}\sigma}$, which recover the conventional
heat current operators \cite{Hardy1963PR}, where ${\bf v}_{{\bf k}\sigma}$
is the phonon group velocity. 

Substituting Eqs. (\ref{eq_HCDOD}) and (\ref{eq_HCDOOD}) into Eq. (\ref{eq_Kubo}), one can
obtain the mean PAM per unit volume as $\<{\mathcal{L}}_{\mu}(\omega)\>=\<{\mathcal{L}}^{\text{D}}_{\mu}(\omega)\>+\<{\mathcal{L}}^{\text{OD}}_{\mu}(\omega)\>$. The diagonal part
$\<{\mathcal{L}}_{\mu}^{\text D}(\omega)\>=\<L_{\mu}^{\text D}(\omega)\>/V$ and the off-diagonal part  $\<{\mathcal{L}}_{\mu}^{\text
  {OD}}(\omega)\>=\<L_{\mu}^{\text {OD}}(\omega)\>/V$ are written as follows, respectively,
  \begin{widetext}
\begin{subequations}
\begin{equation}
\<{\mathcal{L}}_{\mu}^{\text D}(\omega)\>=\sum_{\nu}\frac{i (\nabla
  T)_{\nu}}{T} \frac{1}{\omega+i\eta}\sum\limits_{{\bf
  	k}\sigma}\frac{\partial f^{0}_{{\bf k}\sigma}}{\partial
  (\hbar\omega_{{\bf k}\sigma})}({\bf s}_{{\bf
  	k}\sigma})_{\nu}l^{\mu}_{{\bf k\sigma}}
=\sum_{\nu}\Lambda_{\mu\nu}^{\text{D}}(\omega)\frac{\partial T}{\partial x_{\nu}},
\label{eq_MeanLZD}
\end{equation}

\begin{equation}
\<{\mathcal{L}}_{\mu}^{\rm OD}(\omega)\>=-\sum_{\nu}\frac{i (\nabla
  T)_{\nu}}{T} \sum\limits_{{\bf
    k}\sigma\sigma^{\prime}}\frac{f^{0}_{{\bf k}\sigma}-f^{0}_{{\bf
      k}\sigma^{\prime}}}{\hbar(\omega_{{\bf k}\sigma}-\omega_{{\bf
      k}\sigma^{\prime}})}\frac{({\bf s}_{{\bf k}\sigma \sigma^{\prime}})_{\nu} l^{\mu}_{{\bf k}\sigma^{\prime}\sigma}}{\omega_{\bf k\sigma}-\omega_{\bf k\sigma^{\prime}}-\omega-i\eta}
=\sum_{\nu}\Lambda_{\mu\nu}^{\text{OD}}(\omega)\frac{\partial T}{\partial x_{\nu}},
\label{eq_MeanLZOD}
\end{equation}
The detailed derivations are shown in Appendix A.

%\begin{equation}
%\<L_{\mu}(\omega)\>=\sum_{\nu}(\Lambda_{\mu\nu}^{\text{D}}(\omega)+\Lambda_{\mu\nu}^{\text{OD}}(\omega))\frac{\partial T}{\partial x_{\nu}}
%=\sum_{\nu}\Lambda_{\mu\nu}(\omega)\frac{\partial T}{\partial x_{\nu}}.
%\label{eq_MeanL}
%\end{equation}
\end{subequations}
\end{widetext}

\par
We now numerically calculate the response tensor $\Lambda_{\mu\nu}$ in a dice lattice model \cite{2011Nearly} as an example. We are interested in the dc current in the $\omega \rightarrow 0$ limit and $\eta$ is simply approximate as the phonon relaxation rate $1/\tau$ \cite{Allen1993PRB}. The detailed parameters of material in calculation are given in Supporting Material.
 
\begin{figure}[H] 
\centering
\includegraphics[width=0.8\linewidth]{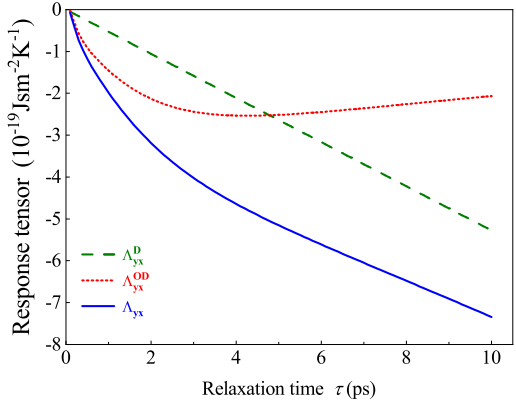}
\caption{(color online) The calculated response tensor $\Lambda_{yx}$ (blue solid line), $\Lambda_{yx}^{\text D}$ (green dashed line), and $\Lambda_{yx}^{\text O\text D}$ (red dotted line) as functions of $\tau$ when $T$ = 300 K.}
\label{fig:1}
\end{figure}

The symmetry of the lattice we studied results in nonzero $xy$ and $yx$ components of the response tensor. 
We first calculate the $\Lambda_{yx}$ at $T$ = 300 K with phonon relaxation time $\tau$ ranging from 0.1 ps to 10 ps as shown in Fig. \ref{fig:1}. As for the diagonal part,  $\Lambda_{yx}^{\text D}$ varies linearly with $\tau$, which is the same as the Boltzmann transport equation calculations \cite{Hamada2018PRL}. In our calculations, we find that $\Lambda_{yx}^{\text D}\sim$ $-10^{-8}$ $\times$ [$\tau$/(1 s)] J s ${\text m}^{-2}$ ${\text K}^{-1}$. More interestingly, we find that the contribution from the off-diagonal part $\Lambda_{yx}^{\text {OD}}$ is also important, especially when $\tau < 5$ ps.
One can see that $\Lambda^{\text{OD}}_{yx}$ decreases first and then increases with increasing $\tau$. The minimum value of $\Lambda^{\text{OD}}_{yx}$ is determined by $\eta\sim \omega_{{\bf k}\sigma}-\omega_{{\bf k}\sigma^{\prime}}$.

\par 

\begin{figure}[H]
\centering
\includegraphics[width=1\linewidth]{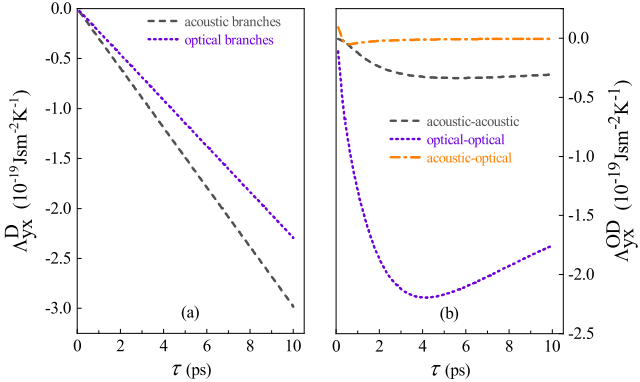}
\caption{(color online) (a) Contribution to $\Lambda_{yx}^{\text {D}}$ from different phonon branches. (b) Contribution to $\Lambda_{yx}^{\text {OD}}$ from transition between different phonon branches.}
\label{fig:2}
\end{figure}
Figure \ref{fig:2} (a) shows the contributions from different phonon branches to $\Lambda_{yx}^{\text D}$.
The calculation results show that the contribution from acoustic phonon is more significant than the contribution from optical phonon. The reason is that the acoustic phonon has larger phonon group velocity which is included in $s_{{\bf k}\sigma}$ in Eq. (\ref{eq_MeanLZD}). In order to distinguish the contribution from the quantum transitions between different phonon branches to off-diagonal PAM, we 
plot the calculated acoustic-acoustic transition, acoustic-optical transition, and optical-optical transition components of $\Lambda_{yx}^{\text {OD}}$ in Fig. \ref{fig:2} (b). The calculation results show that the quantum transition between two optical branches is dominant. The contributions from quantum transition between two acoustic branches and that between one acoustic branch and one optical branch are relatively small. This finding implies that more optical phonon branches are helpful to enlarge the mean PAM. Therefore, chiral materials with many atoms in unit cell, such as the chiral hybrid perovskites, is a good candidate to achieve large PAM \cite{Kim2022NM}.

\begin{figure}[H]
\centering
\includegraphics[width=0.8\linewidth]{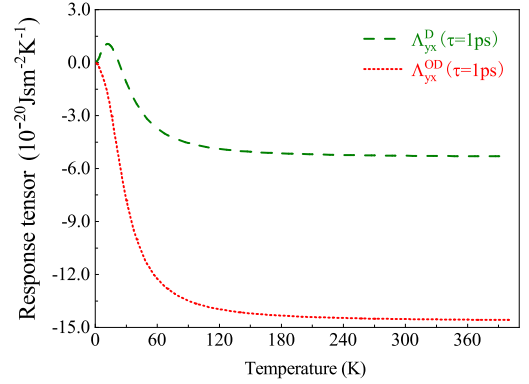}
\caption{(color online). $\Lambda_{yx}^{\text D}$ and $\Lambda_{yx}^{\text {OD}}$ vs temperature $T$ when $\tau=1$ ps.}
\label{fig:3}
\end{figure}

Figure \ref{fig:3} shows the temperature dependence of response tensors. One can see that when $T\rightarrow 0$, both $\Lambda_{yx}^{\text D}$ and $\Lambda_{yx}^{\text {OD}}$ vanish. $\Lambda_{yx}^{\text D}$ at low temperature is positive and increases slowly with increasing $T$. Then $\Lambda_{yx}^{\text D}$ begins a gradual decline and becomes negative. When $T$ is higher than the Debye temperature, $\Lambda_{yx}^{\text D}$ approaches the high temperature limit. The calculation results show that $\Lambda_{yx}^{\text {OD}}$ is more sensitive to temperature. $\Lambda_{yx}^{\text {OD}}$ decreases rapidly with increasing temperature and reaches a high temperature limit. The analytical expressions of response tensor at low-temperature limit and at high-temperature limit as given in Supporting Materials.

In summary, we calculate the phonon angular momentum induced by temperature gradient using the Kubo formula. The off-diagonal elements of density matrix due to the quantum transition between different phonon branches are found to be important to generate the nonzero PAM. Both diagonal and off-diagonal response tensors for certain dice lattice are calculated numerically as an example. Chiral materials with many atoms in unit cell is find to be helpful to observe large PAM and large phonon magnetic moment. In addition, we point out that the off-diagonal PAM could be further enhanced in non-perfect crystals in which the momentum conservation is absent, in other words, the transition between $({\bf k},\sigma)$ and $({\bf k}^{\prime},\sigma^{\prime})$ should be also considered. It will be considered in our future work. To our best knowledge, chiral organic-inorganic hybrid perovskites is a possible candidate to observe large phonon PAM and phonon magnetic moment.  
\par
\section{Acknowledgments}
This work is supported by National Natural Science Foundation of China
(No. 11890703). JZ is also supported by  the “Shuangchuang” Doctor program of Jiangsu Province (JSSCBS20210341).

\section{Appendix A: Kubo formula for PAM}

\setcounter{equation}{0}
\renewcommand{\theequation}{A\arabic{equation}}

The thermal Kubo formula \cite{kubo1957statisticalT} is different from the electrical Kubo formula\cite{kubo1957statisticalE}. 
The reason is that there is a well-defined external force which drives the electrical current. 
However, there is no similar term involving the temperature gradient in the Hamiltonian to drive a heat current.
Thus the thermal Kubo formula requires an additional statistical
hypothesis \cite{zwanzig1965time,mori1962j}, which assumes a local space-dependent temperature $T(x)=[\kB
  \beta(x)]^{-1}$. Then the local density matrix is  
\begin{equation}
	\rho=\dfrac{e^{-\int d^{3}x\beta(x)h(x)}}{Z},
	\label{Aeq1}
\end{equation}
where $h(x)$ is the Hamiltonian density operator, $Z$ is the partition
function, and the Hamiltonian $H=\int d^{3}xh(x)$. 
%In harmonic approximation, the Hamiltonian is
%\begin{equation}
%	H=\sum\limits_{l,\kappa}\dfrac{p^{2}_{l\kappa}}{2m_{\kappa}}+\sum\limits_{l,m}\sum\limits_{\kappa,\kappa^{\prime}}\frac{1}{2}\dfrac{\partial^{2} E}{\partial \uu_{l\kappa} \partial \uu_{m\kappa^{\prime}}}\uu_{l\kappa}\uu_{m\kappa^{\prime}}.
%	\label{Aeq2}
%\end{equation}
A heat current density operator ${\bf S}(x)$ is now defined by the condition of local energy conservation
\begin{equation}
	\dfrac{\partial h(x)}{\partial t}+\nabla\cdot {\bf S}(x)=0.
	\label{Aeq3}
\end{equation}
The total heat current operator is ${\bf S}=\frac{1}{V}\int d^{3}x{\bf
S}(x)$. If the temperature variation $\delta T(x)$ is weak, $\beta(x)$ can be
written as $\beta[1-\delta T(x)]/T$, where $(\kB \beta)^{-1}$ is the
average temperature $T$. Then Eq. (\ref{Aeq1}) becomes $\rho=e^{-\beta (H+H^{\prime})}/Z$
where the operator $H^{\prime}=-\frac{1}{T}\int d^{3}x\delta
T(x)h(x)$ formally plays the role of a
perturbation. Using the integrated form of Eq. (\ref{Aeq3}), the
perturbation due to the temperature gradient is
\begin{equation}
	H^{\prime}=-\sum\limits_{\nu}\dfrac{(\nabla T)_{\nu}}{T}\int_{-\infty}^{0}dt\int d^{3}xS_{\nu}(x,t).
	\label{Aeq5}
\end{equation}
The density matrix can be expended in powers of the perturbation by
$e^{-\beta (H+H^{\prime})} = e^{-\beta H}+e^{-\beta
  H}\int^{\beta}_{0}d\lambda e^{\lambda H}H^{\prime}e^{-\lambda
  H}+\cdots=\rho_{0}+\rho_{1}$ with parameter $\lambda$.
It is obvious that ${\text {tr}}\rho_{0} { L_{\mu}}\equiv 0$, and the
nonzero PAM is determined by ${\text {tr}}\rho_{1}
L_{\mu}$ which is 
\begin{eqnarray}
\<\ L_{\mu}\>&=&-V\sum\limits_{\nu}\dfrac{(\nabla T)_{\nu}}{T}
	 \sum\limits_{n}\dfrac{e^{-\beta E_{n}}}{Z}\int_{0}^{\beta}d\lambda\int_{-\infty}^{0}dt\nonumber\\
	&&\times\<e^{\lambda H}S_{\nu}(t)e^{-\lambda H}L_{\mu}(0)\>.
	\label{Aeq6}
\end{eqnarray}
$e^{\lambda H}S
e^{-\lambda H}=S(-i\hbar\lambda)$, $E_{n}$ is the energy of state $({\bf k},\sigma)$ which is noted
as $n$ for simplicity. Then the Kubo formula for PAM in
Eq. (\ref{eq_Kubo}) can be obtained \cite{allen1989thermal,Allen1993PRB}.
%\begin{eqnarray}
%\<L_{z}(\omega)\>&=&-V
%\sum\limits_{\alpha}\dfrac{(\nabla T)_{\alpha}}{T}\sum\limits_{n}\dfrac{e^{-\beta E_{n}}}{Z}\int_{0}^{\beta}d\lambda\int_{0}^{\infty}dte^{i(\omega+i\eta)t}\nonumber\\
%&&\times\<S_{\alpha}(-i\hbar\lambda)L_{z}(t)\>. 
%\label{Aeq7}
%\end{eqnarray}	
$\omega$ represents the frequency of an ac applied temperature gradient. 
This Kubo expression can be written in the Lehman representation
\begin{widetext}
\begin{equation}
\begin{aligned}
	\<L_{\mu}(\omega)\>=&-V\sum\limits_{\nu}\dfrac{(\nabla T)_{\nu}}{T} \sum\limits_{n,m}\dfrac{e^{-\beta E_{n}}}{Z}\int_{0}^{\beta}d\lambda\int_{0}^{\infty}dte^{i(\omega+i\eta)t} \<n|e^{\lambda H}S_{\nu} e^{-\lambda H}|m\>\<m|e^{\frac{iHt}{\hbar}}L_{\mu}e^{\frac{-iHt}{\hbar}}|n\>, \\
	=&V\sum\limits_{\nu}\dfrac{(\nabla T)_{\nu}}{T} \sum\limits_{n,m}\dfrac{e^{-\beta E_{n}}}{Z} \[\dfrac{e^{\beta (E_{n}-E_{m})}-1}{E_{n}-E_{m}}\] \(\dfrac{i\hbar}{E_{n}-E_{m}-\hbar\omega-i\hbar\eta}\) \<n|S_{\nu}|m\> \<m|L_{\mu}|n\>.		
	\label{Aeq8}
\end{aligned}
\end{equation}
\end{widetext}
This expression can be divided into two parts: the contribution from
diagonal terms $\< L_{\mu}^{\text D}(\omega)\>$ when $n=m$ and the contribution from offdiagonal
term $\< L_{\mu}^{\text {OD}}(\omega)\>$ when $n\neq m$.
\begin{widetext}
\begin{equation}
\begin{aligned}
\< L_{\mu}^{\rm OD}(\omega)\>=&V\sum\limits_{\nu}\dfrac{(\nabla T)_{\nu}}{T} \sum\limits_{nm}\dfrac{e^{-\beta E_{n}}}{Z} \[\dfrac{e^{\beta (E_{n}-E_{m})}-1}{E_{n}-E_{m}}\] \(\dfrac{i\hbar}{E_{n}-E_{m}-\hbar\omega-i\hbar\eta}\) \\
&\times\sum\limits_{{\bf k}{\bf
    k}^{\prime}\alpha\alpha^{\prime}\sigma\sigma^{\prime}} \<n|a_{{\bf
    k}^{\prime}\alpha}^{\dagger}a_{{\bf
    k}^{\prime}\alpha^{\prime}}|m\>\<m|a_{{\bf k}\sigma}^{\dagger}
a_{{\bf k}\sigma^{\prime}}|n\>({\bf s}_{{\bf
    k}^{\prime}\alpha\alpha^{\prime}})_{\nu} {l}^{\mu}_{{\bf k}\sigma\sigma^{\prime}}.	
\label{Aeq9}
\end{aligned}
\end{equation}
\end{widetext}
The exact state $|m\>$ is the unique state which couples to $a_{{\bf
    k}\sigma}^{\dagger} a_{{\bf k}\sigma^{\prime}}|n\>$, denoted as
$|n_{{\bf k}\sigma^{\prime}\sigma}\>$. We can use the standard
expressions $\<m|a_{{\bf k}\sigma}^{\dagger} a_{{\bf
    k}\sigma^{\prime}}|n\>=\sqrt{(n_{{\bf k}\sigma}+1)n_{{\bf
      k}\sigma^{\prime}}}\delta_{mn_{{\bf k}\sigma^{\prime}\sigma}}$
and $\<n|a_{{\bf k}^{\prime}\alpha}^{\dagger}a_{{\bf
    k}^{\prime}\alpha^{\prime}}|n_{{\bf
    k}\sigma^{\prime}\sigma}\>=\sqrt{(n_{{\bf k}\sigma}+1)n_{{\bf
      k}\sigma^{\prime}}}\delta_{\alpha^{\prime}\sigma}\delta_{\alpha\sigma^{\prime}}\delta_{{\bf
  k}{\bf k}^{\prime}}$. 
In terms of the normal mode frequencies, $E_{n}-E_{m}$ becomes
$\hbar(\omega_{{\bf k}\sigma^{\prime}}-\omega_{{\bf k}\sigma})$, $\alpha=\sigma^{\prime}$, and $\alpha^{\prime}=\sigma$. 
Now since $\sigma \neq \sigma^{\prime}$, we can write 
$\sum\limits_{n}\frac{e^{-\beta E_{n}}}{Z}n_{{\bf
    k}\sigma^{\prime}}(n_{{\bf k}\sigma}+1)=\<(n_{{\bf
    k}\sigma}+1)n_{{\bf k}\sigma^{\prime}}\>=\<n_{{\bf
    k}\sigma^{\prime}}\>(\<n_{{\bf k}\sigma}\>+1)$. Using the identity
$(e^{\beta\hbar(\omega_{{\bf k}\sigma^{\prime}}-\omega_{{\bf
      k}\sigma})}-1)\<n_{{\bf k}\sigma^{\prime}}\>(\<n_{{\bf
    k}\sigma}\>+1)=\<n_{{\bf k}\sigma}\>-\<n_{{\bf
    k}\sigma^{\prime}}\>=f^{0}_{{\bf k}\sigma}-f^{0}_{{\bf
    k}\sigma^{\prime}}$, $\<{\mathcal{L}}_{\mu}^{\text {OD}}(\omega)\>$ in Eq. (\ref{eq_MeanLZOD}) can be obtained by $\sigma \leftrightarrow \sigma^{\prime}$.

Considering
$\<m|a^{\dagger}_{\kk\sigma}a_{\kk\sigma}|n\>=n_{\kk\sigma}\delta_{mn}$,
the contribution from diagonal term can be written as
\begin{widetext}
\begin{equation}
\begin{aligned}
\<L_{\mu}^{\rm D}(\omega)\>=-V\beta\sum\limits_{\nu}\dfrac{(\nabla
  T)_{\nu}}{T} \(\dfrac{i}{\omega+i\eta}\) \sum\limits_{{\bf k}{\bf
    k}^\prime\sigma\alpha}& \[  \<n_{{\bf k}^{\prime}\alpha}n_{{\bf
    k}\sigma}\>(\ss_{{\bf k}^{\prime}\alpha})_{\nu}l^{\mu}_{{\bf
    k}\sigma} + \frac{1}{2}\<n_{{\bf k}^{\prime}\alpha}\>({\bf
  s}_{{\bf k}^{\prime}\alpha})_{\nu}l^{\mu}_{{\bf k}\sigma} 
 +\frac{1}{2}({\bf s}_{{\bf k}^{\prime}\alpha})_{\nu}\<n_{{\bf
     k}\sigma}\>l^{\mu}_{{\bf k}\sigma}+\frac{1}{4}({\bf
   s}_{{\bf k}^{\prime}\alpha})_{\nu}l^{\mu}_{{\bf k}\sigma}\].	
\label{Aeq11}	
\end{aligned}
\end{equation}
\end{widetext}
According the Wick's theorem, the factor $\<n_{{\bf
    k}^{\prime}\alpha}n_{{\bf k}\sigma}\>$ can be written as $\<n_{{\bf
    k}^{\prime}\alpha}\>\<n_{{\bf k}\sigma}\>$ plus a correction
$\<n_{{\bf k}\sigma}\>(\<n_{{\bf k}\sigma}\>+1)$. Using the condition
of zero average energy current and PAM in equilibrium and the identity
$\<n_{{\bf k}\sigma}\>(\<n_{{\bf k}\sigma}\>+1)=-\kB
T\frac{\partial\<n_{{\bf k}\sigma}\>}{\partial(\hbar\omega_{{\bf k}\sigma})}$,  $\<{\mathcal{L}}_{\mu}^{\text {D}}(\omega)\>$ in Eq. (\ref{eq_MeanLZD}) can be obtained by dropping the last three terms in the bracket in Eq. (\ref{Aeq11}).

\section{Appendix B: Derivation of off-diagonal heat current density
  operator}
\setcounter{equation}{0}
\renewcommand{\theequation}{B\arabic{equation}}
Hardy \cite{Hardy1963PR} has shown that the quadratic terms of the
heat current operator is
%\begin{widetext}
%\begin{equation}
%\begin{aligned}
%	\SS=&\dfrac{1}{2V}\sum\limits_{l,\kappa}\left\{ \dfrac{p_{l\kappa}}{m_{\kappa}} \[ \dfrac{p_{l\kappa}^{2}}{2m_{\kappa}} +V(\RR_{l\kappa})\]+\rm{H.C.}\right\} \\
%	&+ \dfrac{1}{2i\hbar V}\sum\limits_{lm\kappa\kappa^{\prime}}\left\{(\RR_{l\kappa}-\RR_{m\kappa^{\prime}})\[\dfrac{p_{l\kappa}^{2}}{2m_{\kappa}}, V(\RR_{m\kapp%a^{\prime}}) \] +\rm{H.C.}\right\}.
%\label{Aeq13}
%\end{aligned}
%\end{equation}
%\end{widetext}

\begin{eqnarray}
{\bf S}&=&
\dfrac{1}{2V}\sum\limits_{ll^{\prime}}\sum\limits_{\kappa\kappa^{\prime}}
({\bf R}_{l\kappa}-{\bf R}_{l^{\prime}\kappa^{\prime}})\sum\limits_{\alpha}
\left\{\dfrac{p_{l\kappa\alpha}}{m_{\kappa}}\dfrac{1}{i\hbar}[p_{l\kappa\alpha},
  V({\bf R}_{l^{\prime}\kappa^{\prime}})] \right. \nonumber\\
&&\left. +\dfrac{1}{i\hbar}[p_{l\kappa\alpha}, V({\bf R}_{l^{\prime}\kappa^{\prime}})]\dfrac{p_{l\kappa\alpha}}{m_{\kappa}} \right\}.	
\label{Aeq14}
\end{eqnarray}
where $V({\bf R}_{l^{\prime}\kappa^{\prime}})$ is harmonic potential energy
\begin{equation}
V({\bf R}_{l^{\prime}\kappa^{\prime}})=
\dfrac{1}{2}\sum\limits_{l}\sum\limits_{\kappa}
\sum\limits_{\alpha\gamma} \Phi_{\alpha\gamma}(l\kappa,l^{\prime}\kappa^{\prime})u_{l\kappa\alpha}u_{l^{\prime}\kappa\prime\gamma}.
\label{Aeq15} 
\end{equation}
Here $\alpha,\gamma=x,y,z$. $\Phi$ is the force constant matrix. It is
easy to verify that $\frac{1}{i\hbar}[p_{l\kappa\alpha}, V({\bf
    R}_{l^{\prime}\kappa^{\prime}})]=-\frac{1}{2}\sum_{\beta}\Phi_{\alpha\gamma}(l\kappa,l^{\prime}\kappa^{\prime})u_{l^{\prime}\kappa^{\prime}\gamma}$, then
\begin{equation}
{\bf S}= \dfrac{1}{2V}\sum\limits_{ll^{\prime}}\sum\limits_{\kappa\kappa^{\prime}}
\sum\limits_{\alpha\gamma}({\bf R}_{l^{\prime}\kappa^{\prime}}-{\bf R}_{l\kappa})\Phi_{\alpha\gamma}({l\kappa},{l^{\prime}\kappa^{\prime}}) \dfrac{p_{l\kappa\alpha}}{m_{\kappa}}u_{l^{\prime}\kappa^{\prime}\gamma}.	
\label{Aeq16}
\end{equation}
By using the second quantization form of $u_{l^{\prime}\kappa^{\prime}\gamma}$
and $p_{l\kappa\alpha}$, we obtain
%\begin{subequations}
%\begin{equation}
%u_{m\kappa^{\prime}}^{\gamma}=\sum\limits_{\kk,\sigma} \epsilon^{\gamma}_{\kappa^{\prime} \kk\sigma}e^{i(\RR_{m}\cdot \kk-\omega_{\kk\sigma}t)}\sqrt{\dfrac{\hbar}{2m%_{\kappa^{\prime}}\omega_{\kk\sigma}N}}a_{\kk\sigma}+\rm{H.C.},
%\label{Aeq17}
%\end{equation}
%\begin{equation}
%p_{l\kappa}^{\alpha}=-i\sum\limits_{\kk,\sigma} \epsilon^{\alpha}_{\kappa \kk\sigma}e^{i(\RR_{l}\cdot \kk-\omega_{\kk\sigma}t)}\sqrt{\dfrac{m_{\kappa}\omega_{\kk\sig%ma}\hbar}{2N}}a_{\kk\sigma}+\rm{H.C.}
%\label{Aeq18}
%\end{equation}
%\end{subequations}

%\begin{widetext}
\begin{equation}
	\begin{aligned}
		{\bf S}\approx&\dfrac{1}{2V} \dfrac{i\hbar}{2N}
                \sum\limits_{ll^{\prime}}\sum\limits_{\kappa\kappa^{\prime}}
                \sum\limits_{{\bf k}{\bf k}^{\prime}}
                \sum\limits_{\sigma\sigma^{\prime}}
                ({\bf R}_{l^{\prime}\kappa^{\prime}}-{\bf R}_{l\kappa})
                \sum\limits_{\alpha\gamma}\dfrac{\Phi_{\alpha\gamma}
                 ({l\kappa},{l^{\prime}\kappa^{\prime}})}{\sqrt{m_{\kappa}m_{\kappa^{\prime}}}} \\
	&\times \epsilon_{{\bf k}\sigma\kappa\alpha}^{*}\epsilon_{{{\bf k}^{\prime}}\sigma^{\prime}\kappa^{\prime}\gamma}
                \(\sqrt{\dfrac{\omega_{{\bf
                        k}^{\prime}\sigma^{\prime}}}{\omega_{{\bf
                        k}\sigma}}}a_{{\bf
                    k}^{\prime}\sigma^{\prime}}a_{{\bf k}\sigma}^{\dagger}
                +\sqrt{\dfrac{\omega_{{\bf k}\sigma}}{\omega_{{\bf
                        k}^{\prime}\sigma^{\prime}}}}a_{{\bf
                    k}\sigma}^{\dagger}a_{{\bf k}^{\prime}\sigma^{\prime}}
                \)\\
             & \times e^{i({\bf k}^{\prime}\cdot{\bf R}_{l^{\prime}\kappa^{\prime}}-{\bf k}\cdot{\bf R}_{l\kappa})} \times e^{i({\bf k}\cdot {\bf d}_{k}-{\bf k^{\prime}}\cdot {\bf d}_{k^{\prime}})}.
	\end{aligned}
	\label{Aeq19}
\end{equation}
%\end{widetext}
Here ${\bf d}_{i}$ is the equilibrium position of the $i$th atom relative to the center of the unit cell and we have ignored the terms like $aa$, $a^{\dagger}a^{\dagger}$. Using the definition $\widetilde{\text D}_{\alpha\gamma}({\bf k^{\prime}},\kappa\kappa^{\prime})$=$\sum_{l^{\prime}} \Phi_{\alpha\gamma}(l\kappa,l^{\prime}\kappa^{\prime})e^{i{\bf k^{\prime}}\cdot({\bf R}_{l^{\prime}\kappa^{\prime}}-{\bf R}_{l\kappa})}/\sqrt{m_{\kappa}m_{\kappa^{\prime}}}$, 
$\widetilde{\epsilon}_{{{\bf k}^{\prime}}\sigma^{\prime}\kappa^{\prime}\gamma}=\epsilon_{{{\bf k}^{\prime}}\sigma^{\prime}\kappa^{\prime}\gamma}e^{-i{\bf k}^{\prime}\cdot{\bf d}_{k^{\prime}}}$ and $\frac{1}{N} \sum_{l} e^{i{\bf R}_{l\kappa}\cdot (\kk^{\prime}-\kk)}=\delta_{\kk,\kk^{\prime}}$, we rewrite
Eq. (\ref{Aeq19}) as 
%\begin{widetext}
\begin{equation}
	\begin{aligned}
		{\bf S}\approx&\dfrac{\hbar}{4V}
               \sum\limits_{\bf k}
                \sum\limits_{\kappa\kappa^{\prime}}
                 \sum\limits_{\sigma\sigma^{\prime}} \sum\limits_{\alpha\gamma} \widetilde{\epsilon}^{\hspace{0.1cm}*}_{{\bf k}\sigma\kappa\alpha} \frac{\partial {\widetilde{\text D}_{\alpha\gamma}({\bf k},\kappa\kappa^{\prime})}}{\partial \bf k} \widetilde{\epsilon}_{{\bf k}\sigma^{\prime}\kappa^{\prime}\gamma} \\
		&\times\(\sqrt{\dfrac{\omega_{{\bf
                        k}\sigma^{\prime}}}{\omega_{{\bf
                        k}\sigma}}}a_{{\bf k}\sigma^{\prime}}a_{{\bf k}\sigma}^{\dagger}
                +\sqrt{\dfrac{\omega_{{\bf k}\sigma}}{\omega_{{\bf
                        k}\sigma^{\prime}}}}a_{{\bf
                    k}\sigma}^{\dagger}a_{{\bf k}\sigma^{\prime}}
                \).
	\end{aligned}
	\label{Aeq20} 
\end{equation}
%\end{widetext}
Due to the communication relation $[a_{\kk\sigma^{\prime}},
  a_{\kk\sigma}^{\dagger}]=\delta_{\sigma\sigma^{\prime}}$, we finally
get the diagonal term in Eq. (\ref{eq_HCDOD}) and off-diagonal term in
Eq. (\ref{eq_HCDOOD})
%\begin{widetext}
%\begin{equation}
%	\begin{aligned}
%		\SS^{\rm OD}=\sum\limits_{\kk\sigma\sigma^{\prime}}\ss_{\kk\sigma\sigma^{\prime}}a_{\kk\sigma}^{\dagger}a_{\kk\sigma^{\prime}}e^{i(\omega_{\kk\sigma}-\omega_{\kk\sigma^{\prime}})t}
%	\end{aligned}
%	\label{Aeq21} 
%\end{equation}
%\end{widetext}
%And the diagonal of $\SS$,
%\begin{equation}
%	\SS^{\rm D}=\sum\limits_{\kk\sigma}\ss_{\kk\sigma}\[a_{\kk\sigma}^{\dagger}a_{\kk\sigma} +\frac{1}{2}\],
%	\label{Aeq22}
%\end{equation}	
%where 

%\begin{thebibliography}{}
\normalem

\bibliography{ref}

%apsrev4-2.bst 2019-01-14 (MD) hand-edited version of apsrev4-1.bst
%Control: key (0)
%Control: author (72) initials jnrlst
%Control: editor formatted (1) identically to author
%Control: production of article title (-1) disabled
%Control: page (0) single
%Control: year (1) truncated
%Control: production of eprint (0) enabled
\begin{thebibliography}{21}%
\makeatletter
\providecommand \@ifxundefined [1]{%
 \@ifx{#1\undefined}
}%
\providecommand \@ifnum [1]{%
 \ifnum #1\expandafter \@firstoftwo
 \else \expandafter \@secondoftwo
 \fi
}%
\providecommand \@ifx [1]{%
 \ifx #1\expandafter \@firstoftwo
 \else \expandafter \@secondoftwo
 \fi
}%
\providecommand \natexlab [1]{#1}%
\providecommand \enquote  [1]{``#1''}%
\providecommand \bibnamefont  [1]{#1}%
\providecommand \bibfnamefont [1]{#1}%
\providecommand \citenamefont [1]{#1}%
\providecommand \href@noop [0]{\@secondoftwo}%
\providecommand \href [0]{\begingroup \@sanitize@url \@href}%
\providecommand \@href[1]{\@@startlink{#1}\@@href}%
\providecommand \@@href[1]{\endgroup#1\@@endlink}%
\providecommand \@sanitize@url [0]{\catcode `\\12\catcode `\$12\catcode
  `\&12\catcode `\#12\catcode `\^12\catcode `\_12\catcode `\%12\relax}%
\providecommand \@@startlink[1]{}%
\providecommand \@@endlink[0]{}%
\providecommand \url  [0]{\begingroup\@sanitize@url \@url }%
\providecommand \@url [1]{\endgroup\@href {#1}{\urlprefix }}%
\providecommand \urlprefix  [0]{URL }%
\providecommand \Eprint [0]{\href }%
\providecommand \doibase [0]{https://doi.org/}%
\providecommand \selectlanguage [0]{\@gobble}%
\providecommand \bibinfo  [0]{\@secondoftwo}%
\providecommand \bibfield  [0]{\@secondoftwo}%
\providecommand \translation [1]{[#1]}%
\providecommand \BibitemOpen [0]{}%
\providecommand \bibitemStop [0]{}%
\providecommand \bibitemNoStop [0]{.\EOS\space}%
\providecommand \EOS [0]{\spacefactor3000\relax}%
\providecommand \BibitemShut  [1]{\csname bibitem#1\endcsname}%
\let\auto@bib@innerbib\@empty
%</preamble>
\bibitem [{\citenamefont {Ashcroft}\ and\ \citenamefont
  {Mermin}(1976)}]{Kittelbook}%
  \BibitemOpen
  \bibfield  {author} {\bibinfo {author} {\bibfnamefont {N.~W.}\ \bibnamefont
  {Ashcroft}}\ and\ \bibinfo {author} {\bibfnamefont {N.~D.}\ \bibnamefont
  {Mermin}},\ }\href@noop {} {\emph {\bibinfo {title} {Solid State Physics}}}\
  (\bibinfo  {publisher} {Thomson Learning, Inc.},\ \bibinfo {year} {1976})\
  p.\ \bibinfo {pages} {453}\BibitemShut {NoStop}%
\bibitem [{\citenamefont {Zhu}\ \emph {et~al.}(2018)\citenamefont {Zhu},
  \citenamefont {Yi}, \citenamefont {Li}, \citenamefont {Xiao}, \citenamefont
  {Zhang}, \citenamefont {Yang}, \citenamefont {Kaindl}, \citenamefont {Li},
  \citenamefont {Wang},\ and\ \citenamefont {Zhang}}]{Zhu2018SC}%
  \BibitemOpen
  \bibfield  {author} {\bibinfo {author} {\bibfnamefont {H.}~\bibnamefont
  {Zhu}}, \bibinfo {author} {\bibfnamefont {J.}~\bibnamefont {Yi}}, \bibinfo
  {author} {\bibfnamefont {M.}~\bibnamefont {Li}}, \bibinfo {author}
  {\bibfnamefont {J.}~\bibnamefont {Xiao}}, \bibinfo {author} {\bibfnamefont
  {L.}~\bibnamefont {Zhang}}, \bibinfo {author} {\bibfnamefont
  {C.}~\bibnamefont {Yang}}, \bibinfo {author} {\bibfnamefont {R.~A.}\
  \bibnamefont {Kaindl}}, \bibinfo {author} {\bibfnamefont {L.}~\bibnamefont
  {Li}}, \bibinfo {author} {\bibfnamefont {Y.}~\bibnamefont {Wang}},\ and\
  \bibinfo {author} {\bibfnamefont {X.}~\bibnamefont {Zhang}},\ }\href@noop {}
  {\bibfield  {journal} {\bibinfo  {journal} {Science}\ }\textbf {\bibinfo
  {volume} {359}},\ \bibinfo {pages} {579} (\bibinfo {year}
  {2018})}\BibitemShut {NoStop}%
\bibitem [{\citenamefont {Zhang}\ and\ \citenamefont
  {Niu}(2014)}]{Zhang2014PRL}%
  \BibitemOpen
  \bibfield  {author} {\bibinfo {author} {\bibfnamefont {L.}~\bibnamefont
  {Zhang}}\ and\ \bibinfo {author} {\bibfnamefont {Q.}~\bibnamefont {Niu}},\
  }\href@noop {} {\bibfield  {journal} {\bibinfo  {journal} {Phys. Rev. Lett.}\
  }\textbf {\bibinfo {volume} {112}},\ \bibinfo {pages} {085503} (\bibinfo
  {year} {2014})}\BibitemShut {NoStop}%
\bibitem [{\citenamefont {Zhang}\ and\ \citenamefont
  {Niu}(2015)}]{Zhang2015PRL}%
  \BibitemOpen
  \bibfield  {author} {\bibinfo {author} {\bibfnamefont {L.}~\bibnamefont
  {Zhang}}\ and\ \bibinfo {author} {\bibfnamefont {Q.}~\bibnamefont {Niu}},\
  }\href@noop {} {\bibfield  {journal} {\bibinfo  {journal} {Phys. Rev. Lett.}\
  }\textbf {\bibinfo {volume} {115}},\ \bibinfo {pages} {115502} (\bibinfo
  {year} {2015})}\BibitemShut {NoStop}%
\bibitem [{\citenamefont {McLellan}(1988)}]{Mclellan1988JPC}%
  \BibitemOpen
  \bibfield  {author} {\bibinfo {author} {\bibfnamefont {A.~G.}\ \bibnamefont
  {McLellan}},\ }\href@noop {} {\bibfield  {journal} {\bibinfo  {journal} {J.
  Phys. C: Solid State Phys.}\ }\textbf {\bibinfo {volume} {21}},\ \bibinfo
  {pages} {1177} (\bibinfo {year} {1988})}\BibitemShut {NoStop}%
\bibitem [{\citenamefont {Hamada}\ \emph {et~al.}(2018)\citenamefont {Hamada},
  \citenamefont {Minamitani}, \citenamefont {Hirayama},\ and\ \citenamefont
  {Murakami}}]{Hamada2018PRL}%
  \BibitemOpen
  \bibfield  {author} {\bibinfo {author} {\bibfnamefont {M.}~\bibnamefont
  {Hamada}}, \bibinfo {author} {\bibfnamefont {E.}~\bibnamefont {Minamitani}},
  \bibinfo {author} {\bibfnamefont {M.}~\bibnamefont {Hirayama}},\ and\
  \bibinfo {author} {\bibfnamefont {S.}~\bibnamefont {Murakami}},\ }\href@noop
  {} {\bibfield  {journal} {\bibinfo  {journal} {Phys. Rev. Lett.}\ }\textbf
  {\bibinfo {volume} {121}},\ \bibinfo {pages} {175301} (\bibinfo {year}
  {2018})}\BibitemShut {NoStop}%
\bibitem [{\citenamefont {Park}\ and\ \citenamefont {Yang}(2020)}]{Park2020NL}%
  \BibitemOpen
  \bibfield  {author} {\bibinfo {author} {\bibfnamefont {S.}~\bibnamefont
  {Park}}\ and\ \bibinfo {author} {\bibfnamefont {B.-J.}\ \bibnamefont
  {Yang}},\ }\href@noop {} {\bibfield  {journal} {\bibinfo  {journal} {Nano
  Lett.}\ }\textbf {\bibinfo {volume} {20}},\ \bibinfo {pages} {7694} (\bibinfo
  {year} {2020})}\BibitemShut {NoStop}%
\bibitem [{\citenamefont {Huang}\ \emph {et~al.}(2020)\citenamefont {Huang},
  \citenamefont {Bloom}, \citenamefont {Ni}, \citenamefont {Georgieva},
  \citenamefont {Marciesky}, \citenamefont {Vetter}, \citenamefont {Liu},
  \citenamefont {Waldeck},\ and\ \citenamefont {Sun}}]{Huang2020ACSNano}%
  \BibitemOpen
  \bibfield  {author} {\bibinfo {author} {\bibfnamefont {Z.}~\bibnamefont
  {Huang}}, \bibinfo {author} {\bibfnamefont {B.~P.}\ \bibnamefont {Bloom}},
  \bibinfo {author} {\bibfnamefont {X.}~\bibnamefont {Ni}}, \bibinfo {author}
  {\bibfnamefont {Z.~N.}\ \bibnamefont {Georgieva}}, \bibinfo {author}
  {\bibfnamefont {M.}~\bibnamefont {Marciesky}}, \bibinfo {author}
  {\bibfnamefont {E.}~\bibnamefont {Vetter}}, \bibinfo {author} {\bibfnamefont
  {F.}~\bibnamefont {Liu}}, \bibinfo {author} {\bibfnamefont {D.~H.}\
  \bibnamefont {Waldeck}},\ and\ \bibinfo {author} {\bibfnamefont
  {D.}~\bibnamefont {Sun}},\ }\href@noop {} {\bibfield  {journal} {\bibinfo
  {journal} {ACS Nano}\ }\textbf {\bibinfo {volume} {14}},\ \bibinfo {pages}
  {10370} (\bibinfo {year} {2020})}\BibitemShut {NoStop}%
\bibitem [{\citenamefont {Ren}\ \emph {et~al.}(2021)\citenamefont {Ren},
  \citenamefont {Xiao}, \citenamefont {Sapaov},\ and\ \citenamefont
  {Niu}}]{Ren2021PRL}%
  \BibitemOpen
  \bibfield  {author} {\bibinfo {author} {\bibfnamefont {Y.}~\bibnamefont
  {Ren}}, \bibinfo {author} {\bibfnamefont {C.}~\bibnamefont {Xiao}}, \bibinfo
  {author} {\bibfnamefont {D.}~\bibnamefont {Sapaov}},\ and\ \bibinfo {author}
  {\bibfnamefont {Q.}~\bibnamefont {Niu}},\ }\href@noop {} {\bibfield
  {journal} {\bibinfo  {journal} {Phys. Rev. Lett.}\ }\textbf {\bibinfo
  {volume} {127}},\ \bibinfo {pages} {186403} (\bibinfo {year}
  {2021})}\BibitemShut {NoStop}%
\bibitem [{\citenamefont {Kim}\ \emph {et~al.}(2022)\citenamefont {Kim},
  \citenamefont {Vetter}, \citenamefont {Yan}, \citenamefont {Yang},
  \citenamefont {Wang}, \citenamefont {Sun}, \citenamefont {Yang},
  \citenamefont {Comstock}, \citenamefont {Li}, \citenamefont {Zhou},
  \citenamefont {Zhang}, \citenamefont {You}, \citenamefont {Sun},\ and\
  \citenamefont {Liu}}]{Kim2022NM}%
  \BibitemOpen
  \bibfield  {author} {\bibinfo {author} {\bibfnamefont {K.}~\bibnamefont
  {Kim}}, \bibinfo {author} {\bibfnamefont {E.}~\bibnamefont {Vetter}},
  \bibinfo {author} {\bibfnamefont {L.}~\bibnamefont {Yan}}, \bibinfo {author}
  {\bibfnamefont {C.}~\bibnamefont {Yang}}, \bibinfo {author} {\bibfnamefont
  {Z.}~\bibnamefont {Wang}}, \bibinfo {author} {\bibfnamefont {R.}~\bibnamefont
  {Sun}}, \bibinfo {author} {\bibfnamefont {Y.}~\bibnamefont {Yang}}, \bibinfo
  {author} {\bibfnamefont {A.}~\bibnamefont {Comstock}}, \bibinfo {author}
  {\bibfnamefont {X.}~\bibnamefont {Li}}, \bibinfo {author} {\bibfnamefont
  {J.}~\bibnamefont {Zhou}}, \bibinfo {author} {\bibfnamefont {L.}~\bibnamefont
  {Zhang}}, \bibinfo {author} {\bibfnamefont {W.}~\bibnamefont {You}}, \bibinfo
  {author} {\bibfnamefont {D.}~\bibnamefont {Sun}},\ and\ \bibinfo {author}
  {\bibfnamefont {J.}~\bibnamefont {Liu}},\ }\href@noop {} {\bibfield
  {journal} {\bibinfo  {journal} {Nat. Mater., in revision}\ } (\bibinfo {year}
  {2022})}\BibitemShut {NoStop}%
\bibitem [{\citenamefont {Cheng}\ \emph {et~al.}(2020)\citenamefont {Cheng},
  \citenamefont {Schumann}, \citenamefont {Wang}, \citenamefont {Zhang},
  \citenamefont {Barbalas}, \citenamefont {Stemmer},\ and\ \citenamefont
  {Armitage}}]{Chen2020NL}%
  \BibitemOpen
  \bibfield  {author} {\bibinfo {author} {\bibfnamefont {B.}~\bibnamefont
  {Cheng}}, \bibinfo {author} {\bibfnamefont {T.}~\bibnamefont {Schumann}},
  \bibinfo {author} {\bibfnamefont {Y.}~\bibnamefont {Wang}}, \bibinfo {author}
  {\bibfnamefont {X.}~\bibnamefont {Zhang}}, \bibinfo {author} {\bibfnamefont
  {D.}~\bibnamefont {Barbalas}}, \bibinfo {author} {\bibfnamefont
  {S.}~\bibnamefont {Stemmer}},\ and\ \bibinfo {author} {\bibfnamefont {N.~P.}\
  \bibnamefont {Armitage}},\ }\href@noop {} {\bibfield  {journal} {\bibinfo
  {journal} {Nano Lett.}\ }\textbf {\bibinfo {volume} {20}},\ \bibinfo {pages}
  {5991} (\bibinfo {year} {2020})}\BibitemShut {NoStop}%
\bibitem [{\citenamefont {Nova}\ \emph {et~al.}(2017)\citenamefont {Nova},
  \citenamefont {Cartella}, \citenamefont {Cantaluppi}, \citenamefont
  {F\"{o}rst}, \citenamefont {Bossini}, \citenamefont {Mikhaylovskiy},
  \citenamefont {Kimel}, \citenamefont {Merlin},\ and\ \citenamefont
  {Cavalleri}}]{Nova2017NP}%
  \BibitemOpen
  \bibfield  {author} {\bibinfo {author} {\bibfnamefont {T.~F.}\ \bibnamefont
  {Nova}}, \bibinfo {author} {\bibfnamefont {A.}~\bibnamefont {Cartella}},
  \bibinfo {author} {\bibfnamefont {A.}~\bibnamefont {Cantaluppi}}, \bibinfo
  {author} {\bibfnamefont {M.}~\bibnamefont {F\"{o}rst}}, \bibinfo {author}
  {\bibfnamefont {D.}~\bibnamefont {Bossini}}, \bibinfo {author} {\bibfnamefont
  {R.~V.}\ \bibnamefont {Mikhaylovskiy}}, \bibinfo {author} {\bibfnamefont
  {A.}~\bibnamefont {Kimel}}, \bibinfo {author} {\bibfnamefont
  {R.}~\bibnamefont {Merlin}},\ and\ \bibinfo {author} {\bibfnamefont
  {A.}~\bibnamefont {Cavalleri}},\ }\href@noop {} {\bibfield  {journal}
  {\bibinfo  {journal} {Nat. Phys.}\ }\textbf {\bibinfo {volume} {13}},\
  \bibinfo {pages} {132} (\bibinfo {year} {2017})}\BibitemShut {NoStop}%
\bibitem [{\citenamefont {Hamada}(2021)}]{HamadaThesis}%
  \BibitemOpen
  \bibfield  {author} {\bibinfo {author} {\bibfnamefont {M.}~\bibnamefont
  {Hamada}},\ }\href@noop {} {\emph {\bibinfo {title} {Ph.D. Thesis}}}\
  (\bibinfo  {publisher} {Tokyo Institute of Technology},\ \bibinfo {year}
  {2021})\BibitemShut {NoStop}%
\bibitem [{\citenamefont {Kubo}\ \emph {et~al.}(1957)\citenamefont {Kubo},
  \citenamefont {Yokota},\ and\ \citenamefont
  {Nakajima}}]{kubo1957statisticalT}%
  \BibitemOpen
  \bibfield  {author} {\bibinfo {author} {\bibfnamefont {R.}~\bibnamefont
  {Kubo}}, \bibinfo {author} {\bibfnamefont {M.}~\bibnamefont {Yokota}},\ and\
  \bibinfo {author} {\bibfnamefont {S.}~\bibnamefont {Nakajima}},\ }\href@noop
  {} {\bibfield  {journal} {\bibinfo  {journal} {J. Phys. Soc. Jpn.}\ }\textbf
  {\bibinfo {volume} {12}},\ \bibinfo {pages} {1203} (\bibinfo {year}
  {1957})}\BibitemShut {NoStop}%
\bibitem [{\citenamefont {Zwanzig}(1965)}]{zwanzig1965time}%
  \BibitemOpen
  \bibfield  {author} {\bibinfo {author} {\bibfnamefont {R.}~\bibnamefont
  {Zwanzig}},\ }\href@noop {} {\bibfield  {journal} {\bibinfo  {journal} {Ann.
  Rev. Phys. Chem.}\ }\textbf {\bibinfo {volume} {16}},\ \bibinfo {pages} {67}
  (\bibinfo {year} {1965})}\BibitemShut {NoStop}%
\bibitem [{\citenamefont {Mori}\ \emph {et~al.}(1962)\citenamefont {Mori},
  \citenamefont {Oppenheim},\ and\ \citenamefont {Ross}}]{mori1962j}%
  \BibitemOpen
  \bibfield  {author} {\bibinfo {author} {\bibfnamefont {H.}~\bibnamefont
  {Mori}}, \bibinfo {author} {\bibfnamefont {I.}~\bibnamefont {Oppenheim}},\
  and\ \bibinfo {author} {\bibfnamefont {J.}~\bibnamefont {Ross}},\ }\href@noop
  {} {\emph {\bibinfo {title} {Studies in Statistical Mechanics}}},\
  Vol.~\bibinfo {volume} {1}\ (\bibinfo  {publisher} {North Holland,
  Amsterdam},\ \bibinfo {year} {1962})\ p.\ \bibinfo {pages} {213}\BibitemShut
  {NoStop}%
\bibitem [{\citenamefont {Kubo}(1957)}]{kubo1957statisticalE}%
  \BibitemOpen
  \bibfield  {author} {\bibinfo {author} {\bibfnamefont {R.}~\bibnamefont
  {Kubo}},\ }\href@noop {} {\bibfield  {journal} {\bibinfo  {journal} {J. Phys.
  Soc. Jpn.}\ }\textbf {\bibinfo {volume} {12}},\ \bibinfo {pages} {570}
  (\bibinfo {year} {1957})}\BibitemShut {NoStop}%
\bibitem [{\citenamefont {Allen}\ and\ \citenamefont
  {Feldman}(1993)}]{Allen1993PRB}%
  \BibitemOpen
  \bibfield  {author} {\bibinfo {author} {\bibfnamefont {P.~B.}\ \bibnamefont
  {Allen}}\ and\ \bibinfo {author} {\bibfnamefont {J.~L.}\ \bibnamefont
  {Feldman}},\ }\href@noop {} {\bibfield  {journal} {\bibinfo  {journal} {Phys.
  Rev. B}\ }\textbf {\bibinfo {volume} {48}},\ \bibinfo {pages} {12581}
  (\bibinfo {year} {1993})}\BibitemShut {NoStop}%
\bibitem [{\citenamefont {Hardy}(1963)}]{Hardy1963PR}%
  \BibitemOpen
  \bibfield  {author} {\bibinfo {author} {\bibfnamefont {R.~J.}\ \bibnamefont
  {Hardy}},\ }\href@noop {} {\bibfield  {journal} {\bibinfo  {journal} {Phys.
  Rev.}\ }\textbf {\bibinfo {volume} {132}},\ \bibinfo {pages} {168} (\bibinfo
  {year} {1963})}\BibitemShut {NoStop}%
\bibitem [{\citenamefont {Wang}\ and\ \citenamefont {Ran}(2011)}]{2011Nearly}%
  \BibitemOpen
  \bibfield  {author} {\bibinfo {author} {\bibfnamefont {F.}~\bibnamefont
  {Wang}}\ and\ \bibinfo {author} {\bibfnamefont {Y.}~\bibnamefont {Ran}},\
  }\href@noop {} {\bibfield  {journal} {\bibinfo  {journal} {Phys.rev.b}\
  }\textbf {\bibinfo {volume} {84}},\ \bibinfo {pages} {277} (\bibinfo {year}
  {2011})}\BibitemShut {NoStop}%
\bibitem [{\citenamefont {Allen}\ and\ \citenamefont
  {Feldman}(1989)}]{allen1989thermal}%
  \BibitemOpen
  \bibfield  {author} {\bibinfo {author} {\bibfnamefont {P.~B.}\ \bibnamefont
  {Allen}}\ and\ \bibinfo {author} {\bibfnamefont {J.~L.}\ \bibnamefont
  {Feldman}},\ }\href@noop {} {\bibfield  {journal} {\bibinfo  {journal} {Phys.
  Rev. Lett.}\ }\textbf {\bibinfo {volume} {62}},\ \bibinfo {pages} {645}
  (\bibinfo {year} {1989})}\BibitemShut {NoStop}%
\end{thebibliography}%
%\end{thebibliography}
\end{document}

\title{Supplemental Material for “Abnormal Phonon Angular Momentum due to Off-diagonal Elements in Density Matrix induced by Temperature Gradient}
\author{Jinxin Zhong}
\thanks{These authors contribute equally to this work.}
\affiliation{School of Physics Science and Engineering, Tongji University, 
Shanghai 200092, China}

\author{Hong Sun}
\thanks{These authors contribute equally to this work.}
\affiliation{Phonon Engineering Research Center of Jiangsu Province,
  Center for Quantum Transport and Thermal Energy Science, Institute
  of Physics and Interdisciplinary Science, School of Physics and Technology, Nanjing Normal
	University, Nanjing 210023, China}

\author{Yang Pan}
\affiliation{Phonon Engineering Research Center of Jiangsu Province,
  Center for Quantum Transport and Thermal Energy Science, Institute
  of Physics and Interdisciplinary Science, School of Physics and Technology, Nanjing Normal
	University, Nanjing 210023, China}

\author{Zhiguo Wang}
\affiliation{School of Physics Science and Engineering, Tongji University, 
Shanghai 200092, China}

\author{Xiangfan Xu}
\affiliation{School of Physics Science and Engineering, Tongji University, 
Shanghai 200092, China}

\author{Lifa Zhang}
\email{phyzlf@njnu.edu.cn}
\affiliation{Phonon Engineering Research Center of Jiangsu Province,
  Center for Quantum Transport and Thermal Energy Science, Institute
  of Physics and Interdisciplinary Science, School of Physics and Technology, Nanjing Normal
	University, Nanjing 210023, China}

\author{Jun Zhou}
\email{zhoujunzhou@njnu.edu.cn}
\affiliation{Phonon Engineering Research Center of Jiangsu Province,
  Center for Quantum Transport and Thermal Energy Science, Institute
  of Physics and Interdisciplinary Science, School of Physics and Technology, Nanjing Normal
	University, Nanjing 210023, China}

\date{\today}
\maketitle
\section{{\romannumeral1}.Numerical calculation details}
\begin{figure}[H]
\includegraphics[width=0.6\linewidth]{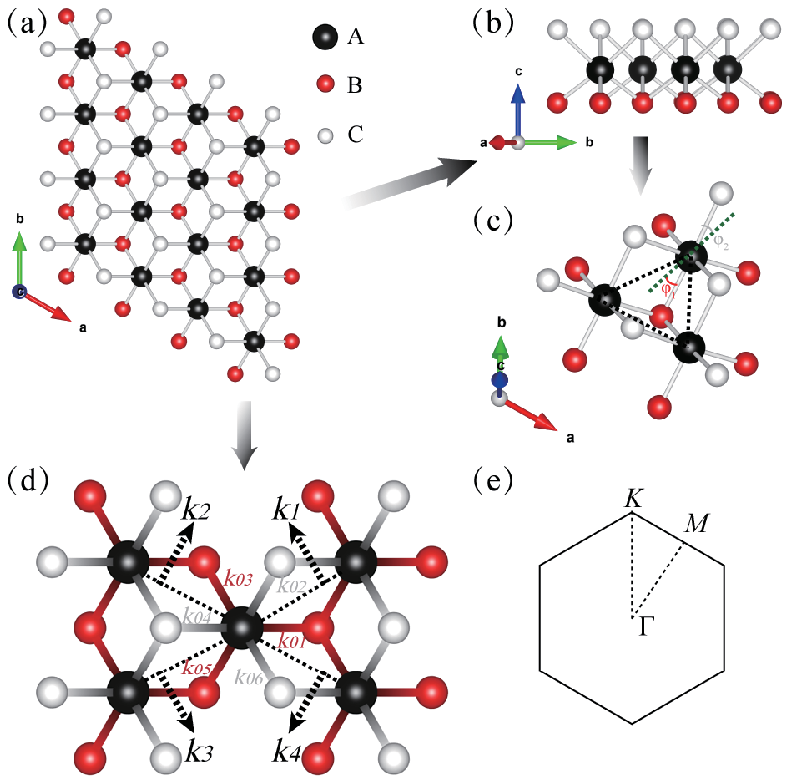}
\centering
\caption{(a) Top view of the dice lattice. The black (middle layer), red (bottom layer) and white (top layer) balls indicate the three sublattices and we label them individually with A, B and C. (b) Side view of the model. It shows a three-level structure. (c) Perspective view of the three adjacent layers. $\varphi_{1}$ is the tilt angel of the bond between atom A and atom B from the $x$-$y$ plane. $\varphi_{2}$ is the tilt angel of the bond between atom A and atom C from the $x$-$y$ plane. (d) A schematic picture of the dice lattice. Each unit cell has three atoms A, B and C. The couplings between the atoms are $K_{01}$, $K_{02}$, $K_{03}$, $K_{04}$, $K_{05}$ and $K_{06}$. Each unit cell has four nearest neighbors; the couplings between the unit cell and the neighbors are $K_1$, $K_2$, $K_3$, and $K_4$. (e) Brillouin zone. $K$ ($k_x$ = 0, $k_y$ = (4$\pi$/3$\sqrt3$$a$)), $M$ ($k_x$ = ($\pi$/3$a$), $k_y$ = ($\sqrt3$$\pi$/3$a$)).}
\label{fig:1}
\end{figure}

In our calculation, a dice lattice model which is shown in Fig. \ref{fig:1} is investigated. 

Owing to the point-group symmetry $C_{3v}$ of the model, the response tensor $\Lambda$ should satisfy $C_{3} \Lambda C_{3}^{-1} = \Lambda$ and $-\sigma_{y} \Lambda \sigma_{y}^{-1} = \Lambda$, where
\begin{equation}
\begin{gathered}
C_{3}=\begin{pmatrix}
-\frac{1}{2}&-\frac{\sqrt3}{2}&0\\
\frac{\sqrt3}{2}&-\frac{1}{2}&0\\
0&0&1\\
\end{pmatrix},
\sigma_{y}=\begin{pmatrix}
1&0&0\\
0&-1&0\\
0&0&1\\
\end{pmatrix}.
\end{gathered}
\end{equation}
One can easily get that only $\Lambda_{yx}$ and $\Lambda_{xy}$ are nonzero.

We first obtain the the dynamical matrix $\text D(\bf{k})$ for the two-dimensional dice lattice. As shown in Fig. \ref{fig:1} (d), each unit cell contains three atoms and only the nearest neighbor interactions are considered. Along $x$-direction, the force constant matrix is set as 
\begin{equation}
{K_{x}}=\begin{pmatrix}
{K_{L}} & 0 & 0\\
0 & {K_{T1}} & 0\\
0 & 0 & {K_{T2}}\\
\end{pmatrix}.\nonumber
\end{equation}
Here, ${K_L}$ is the longitudinal force constant, ${K_{T1}}$ and ${K_{T2}}$ are the transverse force constants. Along $y$-axis, we define a rotation operator $T$($\varphi$) as
\begin{equation}
T({\varphi})=
\begin{pmatrix}
cos\varphi & 0 & -sin\varphi\\
0 & 1 & 0\\
sin\varphi & 0 & cos\varphi\\
\end{pmatrix},\nonumber
\end{equation}
so we can get $K_{x1}$ = $T(-\varphi_{1})$$K_{x}$$T(\varphi_{1})$ and $K_{x2}$ = $T(\varphi_{2})$$K_{x}$$T(-\varphi_{2})$. Along the $z$-axis, we define the rotation operator $U$($\theta$) as
\begin{equation}
U({\theta})=\begin{pmatrix}
cos\theta & sin\theta & 0\\
-sin\theta & cos\theta & 0\\
0 & 0 & 1\\
\end{pmatrix},\nonumber
\end{equation}
the spring-constant matrices between atoms illustrated in Fig. \ref{fig:1} (d) can be written as
\begin{equation}
\begin{aligned}
&k_{01}=K_{x1},\hspace{3 cm}k_{03}=U(-2\pi/3)K_{x1}U(2\pi/3),\\
&k_{05}=U(-4\pi/3)K_{x1}U(4\pi/3),\hspace{0.26cm}k_{02}=U(-\pi/3)K_{x2}U(\pi/3), \\
&k_{04}=U(-\pi)K_{x2}U(\pi),\hspace{1.3cm}k_{06}=U(-5\pi/3)K_{x2}U(5\pi/3), \nonumber
\end{aligned}
\end{equation}
and the on-site spring-constant matrices are defined as
\begin{equation}
\begin{aligned}
k_{11} = k_{01}&+k_{02}+k_{03}+k_{04}+k_{05}+k_{06}, \nonumber\\
&k_{22} = k_{01}+k_{03}+k_{05}, \nonumber\\
&k_{33} = k_{02}+k_{04}+k_{06}. \nonumber\\
\end{aligned}
\end{equation}
\par
Further, we obtain the spring-constant matrices between unit cells as
\begin{equation}
K_0=\begin{pmatrix}
\frac{k_{11}}{m_{\text A}}&\frac{-k_{01}}{\sqrt{m_{\text A}m_{\text B}}}&\frac{-k_{04}}{\sqrt{m_{\text A}m_{\text B}}}\\
\frac{-k_{01}}{\sqrt{m_{\text B}m_{\text A}}}&\frac{k_{22}}{m_{\text B}}&0 \\
\frac{-k_{04}}{\sqrt{m_{\text C}m_{\text A}}}&0&\frac{k_{33}}{m_{\text C}} \\
\end{pmatrix},\nonumber
\end{equation}
\begin{equation}
\begin{gathered}
K_1=\begin{pmatrix}
0&0&\frac{-k_{02}}{\sqrt{m_{\text A}m_{\text C}}}\\
\frac{-k_{05}}{\sqrt{m_{\text B}m_{\text A}}}&0&0\\
0&0&0\\
\end{pmatrix},\nonumber
\quad
K_2=\begin{pmatrix}
0&\frac{-k_{03}}{\sqrt{m_{\text A}m_{\text B}}}&0\\
0&0&0 \\
\frac{-k_{06}}{\sqrt{m_{\text C}m_{\text A}}}&0&0\\
\end{pmatrix},
\end{gathered}
\end{equation}
\begin{equation}
\begin{gathered}
K_3=\begin{pmatrix}
0&\frac{-k_{05}}{\sqrt{m_{\text A}m_{\text B}}}&0\\
0&0&0\\
\frac{-K_{02}}{\sqrt{m_{\text A}m_{\text C}}}&0&0\\
\end{pmatrix},
\quad
K_4=\begin{pmatrix}
0&0&\frac{-k_{06}}{\sqrt{m_{\text A}m_{\text C}}} \\
\frac{-k_{03}}{\sqrt{m_{\text B}m_{\text A}}}&0&0 \\
0&0&0\\
\end{pmatrix}.
\end{gathered}\nonumber
\end{equation}
Finally, the 9$\times$9 dynamic matrix $\text D$$(\bf k)$ can be written as
\begin{equation}
\begin{aligned}
\text D({\bf k})=&K_0+K_1e^{i\frac{3k_{x}+\sqrt3k_{y}}{2}a}+K_2e^{i\frac{-3k_{x}+\sqrt{3}k_{y}}{2}a}+K_3e^{-i\frac{3k_{x}+\sqrt{3}k_{y}}{2}a}+K_4e^{i\frac{3k_{x}-\sqrt{3}k_{y}}{2}a}.
\end{aligned}
\label{eq13}
\end{equation}

The parameters used in the calculations are as follows: $K_{L}$ = 1.44 eV$\mathring{\text A}^{-2}$, $K_{T1}$ = 0.25$K_{L}$, $K_{T2}$ = 0.14$K_{L}$, $\varphi_{1}$ = 0.6 rad, $\varphi_{2}$ = 0.7 rad, the lattice constant $a$ = 10 $\mathring{\text A}$, and the mass of atoms $m_{\text A}$ = 209 u, $m_{\text B}$ = 127 u and $m_{\text C}$ = 126 u. This choice gives longitudinal and transverse sound velocities of 837, 749 and 579 m/s. The temperature gradient is set in the $x$-direction. 
\begin{figure}[H]
\centering
\subfigure{
\includegraphics[width=0.7\linewidth]{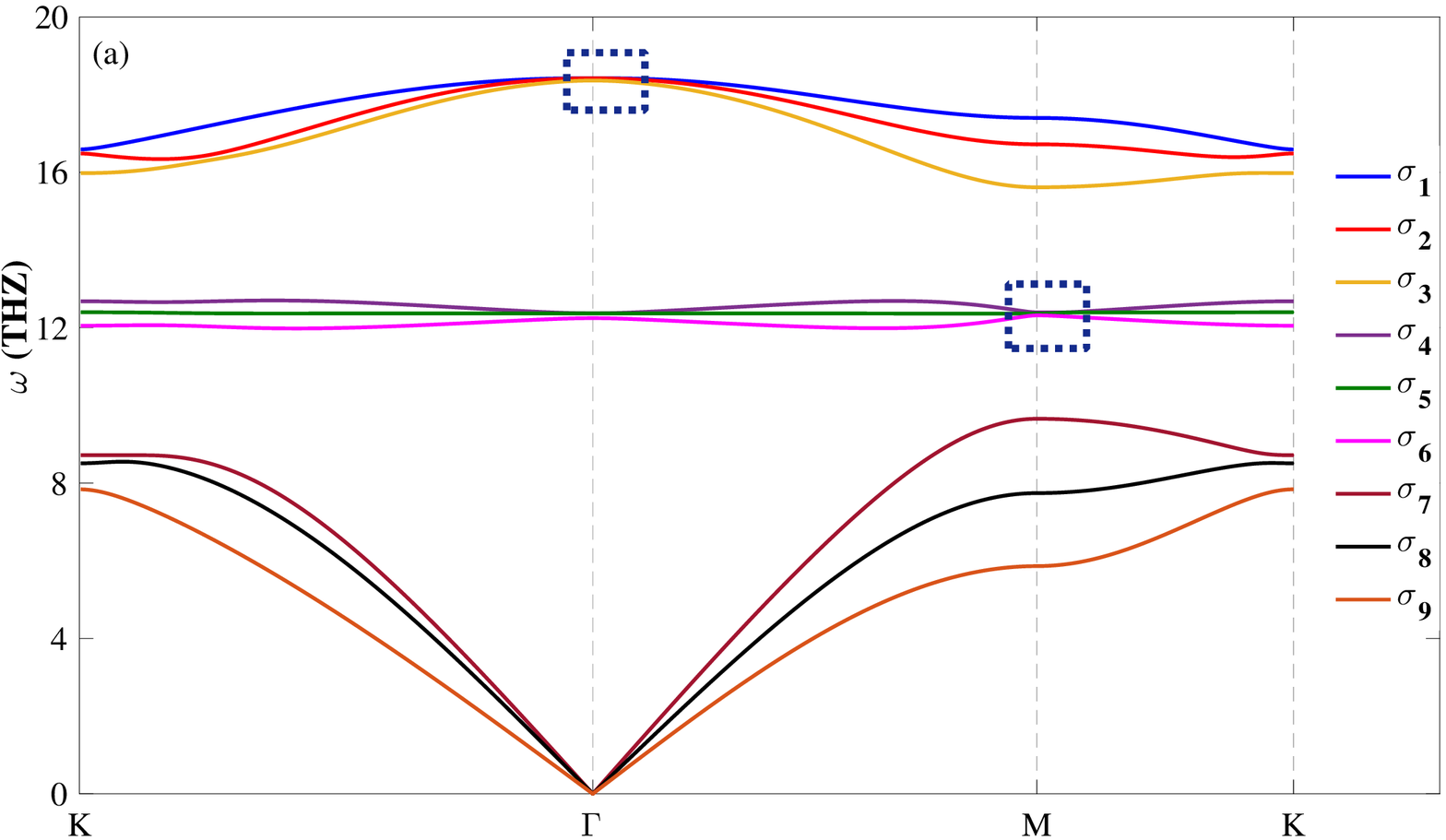}}
\quad
\subfigure{
\includegraphics[width=0.7\linewidth]{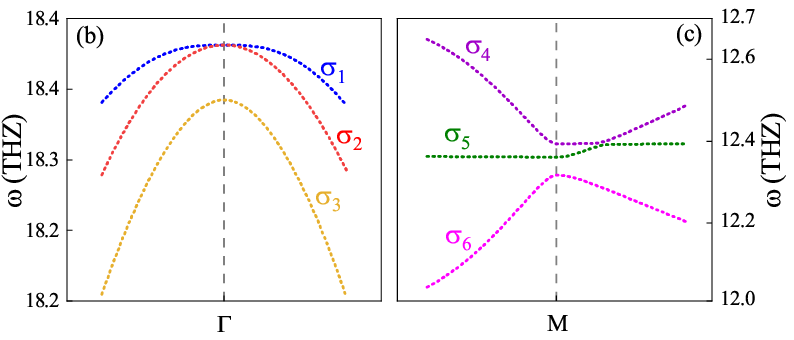}
}
\caption{(a) Dispersion relation along the direction $\text{K}-\Gamma-\text{M}-\text{K}$. (b)-(c) Dispersion diagrams at $\Gamma$ and $\text M$.}
\label{fig:3}
\end{figure}

\section{\expandafter{\romannumeral2}.High- and Low-temperature limits of Response tensor}
In this section, we discuss the temperature dependence of the PAM induced by the temperature gradient at high temperature limit and at low temperature limit.
\par
We first discuss the high temperature limit. When $\hbar \omega_{\bf{k}\sigma} \ll k_{B}T$, the Bose distribution $f_{\bf{k}\sigma}^{0}$  can be expanded in Taylor series as
\begin{equation}
f^{0}(x)=\frac{1}{x}-\frac{1}{2}+\frac{1}{12}x+O\(x^2\) \nonumber
\end{equation}
with $x=\dfrac{\hbar\omega_{\bf{k}\sigma}}{k_{B}T}$. Then the respose tensor $\Lambda$ in the high-temperature limit is represented as
\begin{subequations}
\begin{equation}
\Lambda_{\mu\nu}^{\text D}=\frac{1}{\eta T}\sum\limits_{{\bf
  	k}\sigma}\frac{\partial f^{0}_{{\bf k}\sigma}}{\partial
  (\hbar\omega_{{\bf k}\sigma})}({\bf s}_{{\bf
  	k}\sigma})_{\nu}l^{\mu}_{{\bf k\sigma}}
 \approx-\frac{k_{B}}{V\eta\hbar}\sum\limits_{{\bf k}\sigma}\frac{(v_{{\bf k}{\sigma}})_{\nu}}{\omega_{{\bf k}{\sigma}}}l^{\mu}_{\bf{k}\sigma},
\end{equation}
\begin{equation}
\begin{aligned}
\Lambda_{\mu\nu}^{\text {OD}}&=-\frac{i}{T} \sum\limits_{{\bf
    k}\sigma\sigma^{\prime}}\frac{f^{0}_{{\bf k}\sigma}-f^{0}_{{\bf
      k}\sigma^{\prime}}}{\hbar(\omega_{{\bf k}\sigma}-\omega_{{\bf
      k}\sigma^{\prime}})}\frac{({\bf s}_{{\bf k}\sigma \sigma^{\prime}})_{\nu} l^{\mu}_{{\bf k}\sigma^{\prime}\sigma}}{\omega_{\bf k\sigma}-\omega_{\bf k\sigma^{\prime}}-i\eta}\\
                             &\approx\frac{ik_{B}}{\hbar^{2}}\sum\limits_{{\bf k}{\sigma}{\sigma^{\prime}}}\frac{1}{\omega_{{\bf k}{\sigma}}\omega_{{\bf k}{\sigma^{\prime}}}}\frac{({\bf s}_{{\bf k}\sigma \sigma^{\prime}})_{\nu} l^{\mu}_{{\bf k}\sigma^{\prime}\sigma}}{\omega_{\bf k\sigma}-\omega_{\bf k\sigma^{\prime}}-i\eta}.
\end{aligned}
\end{equation}
\end{subequations}
This shows that the response tensor becomes constant in the high-temperature limit.
\par
Next, we discuss the low temperature limit. When $T\to0$, acoustic modes domain the major contribution comparing with the other phonon modes. Therefore, only the acoustic modes with a long wavelength are considered and the corresponding dispersion relations are represented as
\begin{equation}
\omega_{\sigma}({\bf k})=v_{\sigma}\sqrt{{k_{x}^{2}}+{k_{y}^{2}}}=v_{\sigma}k,\nonumber
\end{equation}
where we assume that the group velocity $v_{\sigma x}=v_{\sigma y}=v_{\sigma}$.
\par
As for $\Lambda_{\mu\nu}^{\text D}$, it's natural to expand $l_{{\bf k}\sigma}^{\mu}$ as
\begin{equation}
l_{{\bf k}{\sigma}}^{\mu}=\alpha_{\sigma x}^{\mu}k_{x}+\alpha_{\sigma y}^{\mu}k_{y}+O(k^{3}).\nonumber
\end{equation}
Since $l_{{\bf k}{\sigma}}^{\mu}$ satisfies $l_{-{\bf k}{\sigma}}^{\mu}=-l_{{\bf k}{\sigma}}^{\mu}$ in the system with time-reversal symmetry, only the odd powers in the wave vector $\bf k$ are preserved. And in a infinite system, the summation of $\bf{k}$ can be replaced by integration,
\begin{equation}
\sum\limits_{{\bf k}}=\frac{A}{(2\pi)^2}\int_{FBZ}dk_{x}dk_{y}=\frac{A}{(2\pi)^2}\int_{-\pi}^{\pi}d\theta\int_{0}^{k_{c}}kdk,\nonumber
\end{equation} 
where $k_{c}$ is a cutoff wavenumber, $A$ is the area. Then $\Lambda_{yx}^{\text D}$ can be written as
\begin{equation}
\begin{aligned}
\Lambda_{yx}^{\text D}&=\frac{A}{4\pi^2\eta T}\sum\limits_{\sigma}\int_{-\pi}^{\pi}d\theta\int_{0}^{k_{c}}kdk\frac{\partial f^{0}_{{\bf k}\sigma}}{\partial
  (\hbar\omega_{{\bf k}\sigma})}({\bf s}_{{\bf
  	k}\sigma})_{x}l^{y}_{{\bf k\sigma}}\\
&=\frac{A}{4\pi^2V\eta T}\sum\limits_{\sigma}\int_{-\pi}^{\pi}d\theta\int_{0}^{k_{c}}k^{2}dk\frac{\partial f^{0}_{{\bf k}\sigma}}{\partial(\omega_{{\bf k}\sigma})}\omega_{{\bf k}\sigma}v_{\sigma}cos\theta \(\alpha_{\sigma x}^{y}cos\theta+\alpha_{\sigma y}^{y}sin\theta\)\\
&=\frac{A}{4\pi V\eta T}\sum\limits_{\sigma}\frac{\alpha_{\sigma x}^{y}}{v_{\sigma}^{2}}\int_{0}^{\omega_{\sigma c}}\omega_{\sigma}^{3}\frac{\partial f^{0}\(\omega_{\sigma}\)}{\partial \omega_{\sigma}}d\omega_{\sigma}. \nonumber
\end{aligned}
\end{equation}
Here we use the relation $\omega_{\sigma}=v_{\sigma}k$ and $\omega_{\sigma c}=v_{\sigma}k_{c}$. By integration by parts, one can obtained $\Lambda_{yx}^{D}$ as
\begin{equation}
\Lambda_{yx}^{D}=\frac{A}{4\pi V\eta T}\sum\limits_{\sigma}\frac{\alpha_{\sigma x}^{y}}{v_{\sigma}^{2}}\(\omega_{\sigma c}^{3}f^{0}\(\omega_{\sigma c}\)-3\int_{0}^{\omega_{\sigma c}}\omega_{\sigma}^{2}f^{0}\(\omega_{\sigma}\)d\omega_{\sigma}\).\nonumber
\end{equation}
The first term vanishes when $T\to0$ and the second term can be simplified as follows:
\begin{equation}
\begin{aligned}
\Lambda_{yx}^{D}&=-\frac{3A k_{B}^{3}}{4\pi V\eta\hbar^{3}}\sum\limits_{\sigma}\frac{\alpha_{\sigma x}^{y}}{v_{\sigma}^{2}}\(\int_{0}^{\infty}\frac{x^{2}}{e^{x}-1}dx\)T^2\\
&=-\frac{3\times1.202A k_{B}^{3}}{2\pi V\eta\hbar^{3}}\(\sum\limits_{\sigma}\frac{\alpha_{\sigma x}^{y}}{v_{\sigma}^{2}}\)T^2.
\end{aligned}
\end{equation}
The temperature dependence of other components can be explained in the same way and  $\Lambda_{\mu\nu}^{\text D}$ is proportional to $T^{2}$ in the low temperature limit.
\par
As for $\Lambda_{\mu\nu}^{\text {OD}}$, we make a rough discussion. As mentioned above, only phonons near the $\Gamma$ point are considered. So the distribution function  $f_{{\bf k}\sigma^{\prime}}^{0}$  can be represented as
\begin{equation}
 f_{{\bf k}\sigma^{\prime}}^{0}\approx f_{{\bf k}\sigma}^{0}+\frac{\partial f_{{\bf k}\sigma}^{0}}{\partial\omega_{{\bf k}\sigma}}\(\omega_{{\bf k}\sigma^{\prime}}-\omega_{{\bf k}\sigma}\), \nonumber
\end{equation}
and $\Lambda_{\mu\nu}^{\text {OD}}$ is rewritten as
\begin{equation}
\Lambda_{\mu\nu}^{\text {OD}}\approx\frac{1}{\hbar\eta T}
\sum\limits_{{\bf k}{\sigma}}\frac{\partial f_{{\bf k}{\sigma}}^{0}}{\partial \omega_{{\bf k}\sigma}}
\sum\limits_{\sigma^{\prime}}\({\bf s}_{{\bf k}\sigma \sigma^{\prime}}\)_{\nu} l^{\mu}_{{\bf k}\sigma^{\prime}\sigma}. \nonumber
\end{equation}
Here we assume that $\(\omega_{{\bf k}{\sigma}}-\omega_{{\bf k}\sigma^{\prime}}\)\ll \eta$ for simplicity. By replacing $\sum\limits_{\sigma^{\prime}}\({\bf s}_{{\bf k}\sigma \sigma^{\prime}}\)_{\nu} l^{\mu}_{{\bf k}\sigma^{\prime}\sigma}$ with $\dfrac{\hbar}{V}G_{\nu\sigma}^{\mu}\({\bf k}\)$ and using Taylor's expansion, one can  get 
\begin{equation}
G_{\nu\sigma}^{\mu}\({\bf k}\)=G_{\nu\sigma}^{\mu}\(0\)+G_{\nu i\sigma}^{\mu}\(0\)k_{i}+\frac{1}{2}G_{\nu i j\sigma}^{\mu}\(0\)k_{i}k_{j}+O\(k^{3}\),\nonumber
\end{equation}
where $\(i,j\)=\(x,y,z\),G_{\nu i\sigma}^{\mu}=\dfrac{\partial G_{\nu\sigma}^{\mu}}{\partial k_{i}}$ 
and $G_{\nu i j\sigma}^{\mu}=\dfrac{\partial G_{\nu\sigma}^{\mu}}{\partial k_{i}\partial k_{j}}$. 
\par After replace summation by integration again, we can obtained $\Lambda_{yx}^{\text {OD}}$ as
\begin{widetext}
\begin{equation}
\begin{aligned}
\Lambda_{yx}^{\text {OD}}&=\dfrac{A}{4\pi^{2}V\eta T}\sum\limits_{\sigma}\int_{0}^{k_{c}}kdk\int_{-\pi}^{\pi}d\theta\frac{\partial f_{{\bf k}\sigma}^{0}}{\partial \omega_{{\bf k}\sigma}}\{G_{x\sigma}^{y}\(0\)+G_{xx\sigma}^{y}\(0\)kcos\theta+G_{xy\sigma}^{y}\(0\)ksin\theta\\
&+\frac{1}{2}[G_{xxx\sigma}^{y}\(0\)k^{2}cos^{2}\theta+G_{xyy\sigma}^{y}\(0\)k^{2}sin^{2}\theta+G_{xxy\sigma}^{y}\(0\)k^{2}sin2\theta]
\}\\
&=\frac{A}{8\pi V\eta T}\sum\limits_{\sigma}\frac{\[G_{xxx\sigma}^{y}\(0\)+G_{xyy\sigma}^{y}\(0\)\]}{v_{\sigma}^{3}}\int_{0}^{\omega_{\sigma c}}\omega_{\sigma}^{3}\frac{\partial f^{0}\(\omega_{\sigma}\)}{\partial \omega_{\sigma}}d\omega_{\sigma}.\nonumber
\end{aligned}
\end{equation}
\end{widetext}
Through simplification, $\Lambda_{yx}^{\text {OD}}$ can be presented as
\begin{equation}
\Lambda_{yx}^{\text {OD}}=-\frac{3\times1.202A k_{B}^{3}}{4\pi V\eta\hbar^{3}}\(\sum\limits_{\sigma}\frac{\[G_{xxx\sigma}^{y}\(0\)+G_{xyy\sigma}^{y}\(0\)\]}{v_{\sigma}^{3}}\)T^{2}.
\end{equation}
This shows that in the low-temperature limit, $\Lambda_{yx}^{\text {OD}}$ is also proprtional to $T^{2}$ and the other components' temperature dependence can be expressed similarly.
\begin{figure}[H]
\centering
\includegraphics[width=0.7\linewidth]{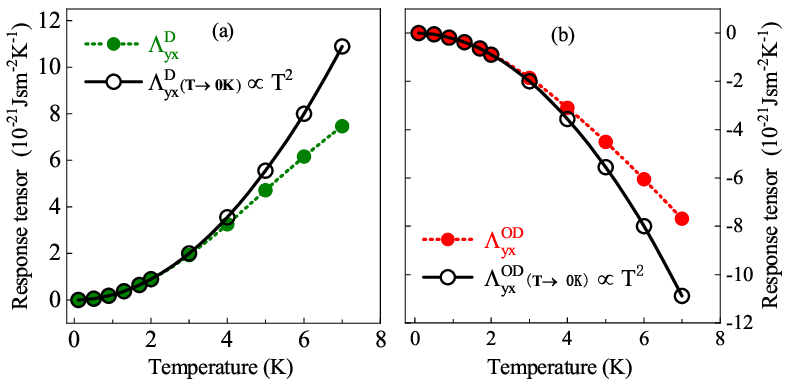}
\caption{(a)-(b) the temperature dependence of $\Lambda_{yx}^{\text D}$ and $\Lambda_{yx}^{\text {OD}}$ in the low temperature limit. The green dotted line (red dotted line) represents the calculational result of $\Lambda_{yx}^{\text D}\(\Lambda_{yx}^{\text {OD}}\)$ from 0.1 K to 7K; The black solid line (empty circle) represents the fitting result}
\label{fig:2}
\end{figure}

%In Figure \ref{fig:2}, we show the corresponding phonon dispersion curve, where the frequency ranges from 0 to 20 THZ. And Figure \ref{fig:2} (a) and (b) represent the phonon dispersion diagrams around the point $\Gamma$ and $\text M$. 